\documentclass{article}

\usepackage[utf8]{inputenc} 
\usepackage[T1]{fontenc}    

\usepackage{url}
\usepackage[pdftex]{hyperref}
\hypersetup{
  pdfauthor={},
  pdftitle={},
  pdfsubject={},
  pdfkeywords={},
  colorlinks=false,
  pdfborder={0 0 0}
}
\usepackage{booktabs}       
\usepackage{amsfonts}       
\usepackage{microtype}      
\usepackage{lipsum}
\usepackage{fancyhdr}       
\usepackage{graphicx}       
\usepackage[title]{appendix}
\graphicspath{{media/}}     
\usepackage{mathtools,amssymb,amsthm,mathrsfs,flexisym,xfrac}
\usepackage{xcolor}
\usepackage{amsmath}
\usepackage{nomencl}
\usepackage{ifthen}
\usepackage[
  a4paper,
  left=2.2cm,
  right=2.2cm,
  top=2.5cm,
  bottom=2.5cm
]{geometry}

\makenomenclature

\title{Optimization of Packed-Bed Energy Storage Systems Based on a Second Law Analysis}

\author{
  Y. Jin$^{*}$, E. Makhova, A. Speerforck \\
  Institute of Engineering Thermodynamics, \\
  Hamburg University of Technology, 21073, Hamburg, Germany \\
  \texttt{Corresponding author: Y. Jin (yan.jin@tuhh.de)} \\ 
}

\begin{document}
\maketitle

\begin{abstract}
Packed-bed sensible heat storage (SHS) is important for balancing energy supply and demand over time. To improve the efficiency of a packed-bed SHS system through second law analysis (SLA), we developed macroscopic entropy and exergy transport equations for fluid flow and heat transfer in porous media based on microscopic transport equations. These equations enable us to identify where and how much exergy is destroyed. Using a packed-bed SHS system developed at the PROMES-CNRS laboratory as a test case, we demonstrated how to apply SLA to optimize an SHS system. Our analysis revealed that, in addition to exit and heat leakage losses at tank surfaces, thermal and solid conduction losses inside the tank significantly contribute to total loss in the studied SHS system. These internal losses occur close to the thermocline. However, their slower transport causes a delay in their emergence. The SLA suggests an optimal tank aspect ratio of $D/H = 0.75$, at which the total exergy loss coefficient, $\zeta^b_{tot}$, reaches its minimum value when exit loss is not considered. As particle size decreases, $\zeta^b_{tot}$ also decreases due to enhanced heat transfer between the fluid and solid phases. The pressure loss for the studied SHS system is negligible. The SLA favors a truncated cone-shaped tank with a slightly larger upper surface. Through the SLA, $\zeta_{tot}^b$ is reduced from $4.9\%$ for the original design to $4.1\%$ for the optimized design. This study demonstrates that, when used in conjunction with energy analysis, the SLA is an effective tool for optimizing energy storage systems.
\end{abstract}


\section{Introduction}
\label{sec:Introduction}

Thermal energy storage (TES) is a method of capturing thermal energy for later use. It allows for a better match between energy supply and demand over time. TES is important for using renewable energies such as solar and wind energy, which are intermittent and not always produced when needed. The development of efficient TES systems is also important for improving energy efficiency, reducing environmental impact, and providing low-cost energy.

There are three main types of TES systems: sensible heat storage (SHS), latent heat storage (LHS), and thermochemical heat storage (TCHS) \cite{Medrano2010,  Zhang2020FractalFins, Gil2010}. Of these, SHS is considered the simplest and most mature. In an SHS system, thermal energy is stored or released by raising or lowering the temperature of thermal energy storage materials (TESMs). These materials can be stored in either one or two tanks \cite{Herrmann2004, Qin2012}. This study focuses on one-tank SHS systems. In these systems, hot fluid is introduced from the top during charging and cold fluid is introduced from the bottom during discharging. Consequently, a thermal gradient, or thermocline, occurs between the hot and cold zones in the tank. Thus, this system is also referred to as a thermocline thermal energy storage (TTES) system \cite{hoffmannThermoclineThermalEnergy2016}.

While liquid TESM is often used in a SHS system, solid materials are sometimes added to the tank to partly replace the liquid TESM, this is called packed-bed one-tank SHS \cite{Xu2012}. This further reduces the cost of the TES, however, the solid particles also make the fluid flow and heat transfer inside the tank more complicated. The relevant parameters that have been studied in the literature include tank height $L$, tank diameter $D$, porosity $\phi$, particle diameter $d_p$, heat transfer fluid (HTF) mass flow rate $\dot{m}$, HTF cold and hot temperatures $T_c$ and $T_h$, TESM properties, etc. \cite{filalibabaDimensionlessModelBased2020}

SHS systems can be studied using experimental or numerical methods. One-dimensional simulations are most often used among the numerical simulations due to their low computational cost. Hoffmann, et al. (2016) \cite{hoffmannThermoclineThermalEnergy2016} argue that a one-dimensional model is sufficient to describe the general behavior of an SHS system. However, some studies use multi-dimensional computational fluid dynamics (CFD) simulations to investigate the flow and temperature fields in more detail. Many CFD studies have been conducted to improve the understanding of thermal stratification during charging and discharging. For instance, Zachar, et al. (2003) examined the impact of various plate sizes positioned opposite the inlet to enhance thermal stratification \cite{Zachar2003}. Göppert, et al. (2009) proposed an efficient CFD method to study the impact of geometry on thermal stratification that considerably reduces computational costs \cite{Goppert2009}.  Kursun (2018) studied the effect of insulation geometries on thermal stratification and proposed a novel geometry to enhance it \cite{Kursun2018}. Li, et al. (2018) found that a truncated circular cone tank has the best temperature stratification and thermal charging efficiency \cite{Li2018}. Gao, et al. (2021) investigated the effect of a baffle plate on thermal stratification characteristics \cite{Gao2021}. Hosseinnia, et al. (2021) simulated thermocline evolution during the charging phase of a stratified thermal energy storage tank. Their numerical results show that an ideal diffuser generates the minimum charging velocity with no mixing or turbulence \cite{Hosseinnia2021}. 

CFD simulations were also used to study the effect of other parameters of SHS systems. For instance, Cascetta, et al. (2016) compared CFD and experimental results from a sensible thermal energy storage system. They indicated that the thermal properties of both phases must be temperature-dependent in the simulation to achieve high accuracy \cite{Cascetta2016}. Khan, et al. (2019) simulated the initial charge-discharge cycle of an SHS system consisting of rocks and pebbles poured into a mild steel tank. They validated the numerical results with experimental data \cite{Khan2019}. Andreozzi, et al. (2014) simulated the charge and discharge cycles of high-temperature thermal storage in a honeycomb solid matrix. They analyzed the effects of the storage medium, different porosity values, and mass flow rate on stored thermal energy and storage time based on the numerical results \cite{Andreozzi2014}.
Elfeky, et al. (2021) investigated the performance of an air-rock thermocline thermal energy storage (TES) tank for concentrated solar power plants using a coupled discrete element method (DEM)-computational fluid dynamics (CFD) approach. Their numerical results show that a HTF distributor can  reduce the heat leakage at the wall \cite{Elfeky2021}. 
Chekifi and Boukraa (2023) conducted a comprehensive review of CFD studies of SHS systems \cite{chekifiCFDApplicationsSensible2023}. They conclude that CFD is a powerful tool for understanding the effects of relevant parameters and optimizing SHS systems.

It can be observed that most numerical studies focus on the storage efficiency based on the the first law of thermodynamics. In particular these studies often only consider the loss of the thermal energy. The corresponding storage efficiency is defined as, 
\begin{equation}
    \eta_I=\frac{Q_{dis}}{Q_{chg}}, 
    \label{eta_I}
\end{equation}
where $Q_{dis}$ is the quantity of thermal energy that is extracted during discharge and $Q_{chg}$ is the thermal energy entering the store during charge \cite{filalibabaDimensionlessModelBased2020}. However, this definition neglects the difference of the energy quality, meaning that the same amount of thermal energy but at different temperatures may produce different amounts of work. To assess the difference in energy quality, one must consider the loss of exergy using the second law of thermodynamics. 

Some works about using the second law analysis (SLA) to study the packed-bed thermal reservoirs exist in the literature. According to the SLA, the efficiency of an SHS system should be evaluated based on the exergy loss, which is defined as  
\begin{equation}
    \eta_{II}=\frac{B_{dis}}{B_{chg}}, 
    \label{eta_II}
\end{equation}
where $B_{dis}$ is the quantity of exergy that is extracted during discharge and $B_{chg}$ is the exergy entering the store during charge. White, et al. (2011) demonstrate the dependence of the losses on operating temperatures, reservoir geometry and mode of operation, and suggest methods of optimization \cite{White2011}. In a later study, White, et al. (2014) suggest that the changes in specific heat capacity result in non-linear wave propagation, which may lead to the formation of shock-like thermal fronts. This has an significant impact on the exergetic losses due to irreversible heat transfer \cite{White2014}. They have analyzed the effect of segmenting the packed-bed into layers on exergy losses \cite{whiteAnalysisOptimisationPackedbed2016, McTigue2016}. The TES system was optimized based on the SLA \cite{McTigue2015}. After optimization, losses due to pressure drop and irreversible heat transfer in the thermal reservoirs are only a few percent. Abdulla and Reddy (2017) suggest that reduction in efficiency at higher temperature difference corresponds to higher value of exergy destruction inside the tank \cite{Abdulla2017ThermoclineTES}. McTigue, et al. (2018) compared the exergy losses of radial-flow and axial-flow packed beds for thermal energy storage, indicating that both designs have comparable thermodynamic performance \cite{McTigue2018b}. They have also found that segmenting the packed bed into layers reduces the effect of perturbations during charge-discharge cycles \cite{McTigue2018}. 

These studies demonstrate the potential of using the SLA to optimize SHS systems. Herwig (2012) \cite{Herwig2012} state that the SLA can be used to answer four important questions concerning a momentum and/or heat transfer process. They are "Which is the ideal process (no entropy generation)? Where does entropy generation occur in a non-ideal process? Why does entropy generation occur at a certain location and with certain strength? How can entropy generation be reduced overall or locally?" The answers to these questions are important for optimizing SHS systems. 
However, to answer the questions, one must perform multi-dimensional (two- or three-dimensional) simulations of the fluid flow and heat transfer problems over a long period of time (typically several hours or even longer), which often requires significant computational resources. Additionally, developing multi-dimensional entropy and exergy transport equations in an SHS system is important for evaluating local entropy generation and exergy destruction intensities. Entropy and exergy transport in an SHS system is more complicated than in pure flow and heat transfer problems because the effect of the packed bed (modeled as a porous medium) must be considered. 

The present study aims to develop multi-dimensional transport equations relevant to the SLA of SHS systems. Then, we will analyze a packed-bed storage system with the SLA to demonstrate its application. The fundamental questions posed by Herwig (2012) \cite{Herwig2012} will be answered for the SHS system studied. The paper is structured as follows: After the introduction, section 2 introduces the governing equations for multi-dimensional computational fluid dynamics (CFD) simulations of fluid flow and heat transfer problems  in a porous medium. This section also derives the macroscopic entropy and exergy transport equations. Section 3 introduces the test case used to demonstrate the application of the SLA. Section 4 discusses the main results. Section 5 provides the conclusions.
 
\section{Mathematical equations and numerical methods}

\subsection{Microscopic transport equations}
The governing equations for pore-scale resolved microscopic simulations of the charging and discharging processes in a packed-bed energy storage system are the classic equations for mass(\ref{equ::mass}), momentum(\ref{equ:momentum}), and energy conservation in both the fluid(\ref{equ:energy_fluid}) and solid(\ref{equ:energy_solid}) phases. These equations can be expressed as follows:

\begin{equation}
    \frac{\partial\rho_f}{\partial t}+\frac{\partial\left(\rho_f u_i\right)}{\partial x_i}=0,
    \label{equ::mass}
\end{equation}
\begin{equation}
    \rho_f\frac{d u_i}{d t} = 
    \frac{\partial \left(\rho_f u_i\right)}{\partial t}+\frac{\partial \left(\rho_f u_j u_i\right)}{\partial x_j}
    =-\frac{\partial p}{\partial x_i}+\frac{\partial \tau_{ji}}{\partial x_j} +\rho_f g_i,
    \label{equ:momentum}
\end{equation}
\begin{equation}
    \rho_f\frac{d h_f}{d t} = 
    \frac{\partial \left(\rho_f h_f\right)}{\partial t}+\frac{\partial \left(\rho_f u_i h_f\right)}{\partial x_i}
    =\frac{d p}{d t}+\frac{\partial }{\partial x_i}\left(k_f\frac{\partial T_f}{\partial x_i}\right) +\tau_{ji}\frac{\partial u_i}{\partial x_j},
    \label{equ:energy_fluid}
\end{equation}
\begin{equation}
    \rho_s\frac{d e_s}{d t} 
    =\frac{\partial }{\partial x_i}\left(k_s\frac{\partial T_s}{\partial x_i}\right).
    \label{equ:energy_solid}
\end{equation}
$\rho_f$, $u_i$, $\tau_{ji}$, $h_f$ and $k_f$ in the governing equations are the density, velocity component, viscous stress, specific enthalpy and thermal conductivity in the fluid phase, respectively. $\rho_s$, $e_s$ and $k_s$ are the density, energy and thermal conductivity in the solid phase, respectively. Multiplying equation (\ref{equ:momentum}) with $u_i$, the transport equation of the kinetic energy $k=\frac{1}{2}u_i^2$ can be obtained in the fluid phase, expressed as, 
\begin{equation}
    \rho_f\frac{d k}{d t} = 
    \frac{\partial \left(\rho_f k\right)}{\partial t}+\frac{\partial \left(\rho_f u_i k\right)}{\partial x_i}
    =-u_i\frac{\partial p}{\partial x_i}+u_i\frac{\partial \tau_{ji}}{\partial x_j} +\rho_f u_i g_i,
    \label{equ:kinetic_energy}
\end{equation}

For a Newtonian fluid, the viscous stress is calculated as   
\begin{equation}
    \tau_{ij} 
    =2 \mu \left(s_{ij}-\frac{2}{3}\delta_{ij}\frac{\partial u_k}{\partial x_k}\right)
    +\eta \delta_{ij} \frac{\partial u_k}{\partial x_k},
    \label{equ:viscous_stress}
\end{equation}
where $s_{ij}$ is the strain rate tensor. $\mu$ and $\eta$ are the  shear viscosity and bulk viscosity, respectively. For an ideal gas, there is the following equation of state:
\begin{equation}
    p=\rho_f R T_f,
    \label{equ:EOS}
\end{equation}
where R is the specific gas constant. $\rho_f$ is a constant for an incompressible fluid.

Considering the fundamental equation of thermodynamics $dh_f=T_f ds_f+\frac{1}{\rho_f}dp$, there are
\begin{equation}
    \frac{dh_f}{dt}=T_f\frac{ds_f}{dt}+\frac{1}{\rho}\frac{dp}{dt},
    \label{equ:EOT_f}
\end{equation}
for the fluid phase and 
\begin{equation}
    \frac{dh_s}{dt}=T_f\frac{ds_s}{dt},
    \label{equ:EOT_s}
\end{equation}
for the solid phase. Here, $s_f$ and $s_s$ represent the specific entropies of the fluid and solid phases, respectively. By substituting equations (\ref{equ:energy_fluid}) and (\ref{equ:energy_solid}) into equations (\ref{equ:EOT_f}) and (\ref{equ:EOT_s}), we can obtain the following transport equations of entropy in the fluid and solid phases:
\begin{equation}
    \rho_f\frac{d s_f}{d t} = 
    \frac{\partial \left(\rho_f s_f\right)}{\partial t}+\frac{\partial \left(\rho_f u_i s_f\right)}{\partial x_i}
    =\frac{\partial }{\partial x_i}\left(\frac{k_f}{T_f}\frac{\partial T_f}{\partial x_i}\right) +\frac{k_f}{T_f^2}\frac{\partial T_f}{\partial x_i}\frac{\partial T_f}{\partial x_i}+\frac{\tau_{ji}s_{ji}}{T_f},
    \label{equ:entropy_fluid}
\end{equation}
\begin{equation}
    \rho_s\frac{d s_s}{d t} =
    \frac{\partial }{\partial x_i}\left(\frac{k_s}{T_s}\frac{\partial T_s}{\partial x_i}\right) +\frac{k_s}{T_s^2}\frac{\partial T_s}{\partial x_i}\frac{\partial T_s}{\partial x_i}.
    \label{equ:entropy_solid}
\end{equation}
Here, $\dot{s}^{mic}_{htf}=\frac{k_f}{T_f^2}\frac{\partial T_f}{\partial x_i}\frac{\partial T_f}{\partial x_i}$ and $\dot{s}^{mic}_{pre}=\frac{\tau_{ji}s_{ji}}{T_f}$ are the entropy generation intensities in the temperature and flow fields of the fluid phase. $\dot{s}^{mic}_{hts}=\frac{k_s}{T_s^2}\frac{\partial T_s}{\partial x_i}\frac{\partial T_s}{\partial x_i}$ is the entropy generation intensity in the temperature field of the solid phase.

Dividing the exergy difference in the fluid phase, $db_f=dh_f-T_0ds_f+dk$, and in the solid phase $db_s=dh_s-T_0ds_s$ by $dt$, we can obtain
\begin{equation}
    \frac{db_f}{dt}=\frac{dh_f}{dt}-T_0\frac{ds_f}{dt}+\frac{dk}{dt},
    \label{equ:EOT_exergy_f}
\end{equation}
for the fluid phase and 
\begin{equation}
    \frac{db_s}{dt}=\frac{de_s}{dt}-T_0\frac{ds_s}{dt},
    \label{equ:EOT_exergy_s}
\end{equation}
for the solid phase. $b_f$ and $b_s$ are the specific exergy in the fluid and solid phases, respectively. Substituting equations (\ref{equ:energy_fluid}), (\ref{equ:entropy_fluid}) and (\ref{equ:kinetic_energy}) into equation (\ref{equ:EOT_exergy_f}), the following transport equation of the exergy in the fluid phase can be obtained, 
\begin{equation}
    \begin{split}   
    \rho_f\frac{d b_f}{d t} = 
    \frac{\partial \left(\rho_f b_f\right)}{\partial t}+\frac{\partial \left(\rho_f u_i b_f\right)}{\partial x_i}
    =\frac{\partial p}{\partial t}+
    \frac{\partial }{\partial x_i}\left(k_f\left(1-\frac{T_0}{T_f}\right)\frac{\partial T_f}{\partial x_i}\right) 
    +\frac{\partial}{\partial x_j}\left(u_i\tau_{ji}\right)
    -T_0\left(\dot{s}^{mic}_{htf}+\dot{s}^{mic}_{pre}\right).
    \end{split}
    \label{equ:exergy_fluid}
\end{equation}
Similarly, substituting equations (\ref{equ:energy_solid}) and (\ref{equ:entropy_solid}) into equation (\ref{equ:EOT_exergy_s}) leads to the exergy transport equation in the solid phase, 
\begin{equation}
    \rho_s\frac{d b_s}{d t} =
    \frac{\partial }{\partial x_i}\left(k_s\left(1-\frac{T_0}{T_s}\right)\frac{\partial T_s}{\partial x_i}\right) 
    -T_0\dot{s}^{mic}_{hts}.
    \label{equ:exergy_solid}
\end{equation}

\subsection{Macroscopic transport equations}
Taking the volume averaging of the microscopic governing equations (\ref{equ::mass})-(\ref{equ:energy_solid}) in a representative elementary volume (REV) leads to the macroscopic governing equations, expressed as 
\begin{equation}
    \frac{\partial\left(\phi\left<\rho_f\right>^f\right)}{\partial t}+\frac{\partial\left(\phi\left<\rho_f\right>^f \left<u_i\right>^f\right)}{\partial x_i}=0,
    \label{equ::macro_mass}
\end{equation}
\begin{equation}
    \begin{split} 
    \left<\rho_f\frac{d u_i}{dt}\right>=
    \frac{\partial \left(\phi\left<\rho_f\right>^f \left<u_i\right>^f\right)}{\partial t}
    +\frac{\partial \left(\phi \left<\rho_f\right>^f \left<u_j\right>^f \left<u_i\right>^f\right)}{\partial x_j}
    =-\phi\frac{\partial \left<p\right>^f}{\partial x_i}
    +\phi\frac{\partial \left<\tau_{ji}\right>^f}{\partial x_j} +\phi\left<\rho_f\right>^f g_i+\phi F_i,
    \end{split}
    \label{equ:macro_momentum}
\end{equation}
\begin{equation}
\begin{split}
    \left<\rho_f\frac{d h_f}{dt}\right>= \frac{\partial \left(\phi\left<\rho_f\right>^f \left<h_f\right>^f\right)}{\partial t}+\frac{\partial \left(\phi\left<\rho_f\right>^f \left<u_i\right>^f \left<h_f\right>^f\right)}{\partial x_i}
    =\phi\frac{\partial \left<p\right>^f}{\partial t} \\
    + \phi\left<u_i \right>^f \frac{\partial \left<p\right>^f}{\partial x_i}
    +\frac{\partial }{\partial x_i}\left(\Tilde{k}_f\frac{\partial \left<T_f\right>^f}{\partial x_i}\right)  +\phi\left<\tau_{ji}\frac{\partial u_i}{\partial x_j}\right>^f
    +\alpha A_{v}\left(\left<T_s\right>^s-\left<T_f\right>^f\right),
    \end{split}
    \label{equ:macro_energy_fluid}    
\end{equation}
\begin{equation}
    \left<\rho_s\frac{d h_s}{dt}\right>=
    \frac{\partial \left(\left(1-\phi\right)\left<\rho_s\right>^s\left<e_s\right>^s\right)}{\partial t} 
    =\frac{\partial }{\partial x_i}\left(\Tilde{k}_s\frac{\partial \left<T_s\right>^s}{\partial x_i}\right)
    +\alpha A_{v}\left(\left<T_f\right>^f-\left<T_s\right>^s\right).
    \label{equ:macro_energy_solid}
\end{equation}
Here, the operator $\left<\right>$ denotes the volume averaging in an REV. $\left<\right>^f$ and $\left<\right>^s$ denote the volume averaging in the fluid or solid phase of an REV. The dispersion effect due to the coupling of two variables is neglected in these equations, e.g., it is assumed that $\phi\left<\rho_fu_i\right>^f$ is identical to $\phi\left<\rho_f\right>^f \left<u_i\right>^f$. 
$\Tilde{k}_f$ and $\Tilde{k}_s$ in the energy equations (\ref{equ:macro_energy_fluid}) and (\ref{equ:macro_energy_solid})are the effective thermal conductivities in the fluid and solid phases, respectively, which need to be modeled in a simulation. $A_v$ is the volume averaged fluid-solid interface area. $\alpha$ is the heat transfer coefficient. 
$F_i$ is the total drag by the porous matrix, which can be approximated using the Forchheimer extension of the Darcy law, expressed as, 
\begin{equation}
    F_i=\frac{\mu}{K} u_{Di}+\frac{c_F}{\sqrt{K}}\left|\textbf{u}_D\right|u_{Di},
    \label{equ:total_drag}
\end{equation}
where $u_{Di}=\phi\left<u_i\right>^f$ is the superficial velocity. $K$ is the permeability. $c_F$ is the Forchheimer constant. 

Multiplying equation (\ref{equ:momentum}) with $u_i$ and performing the volume averaging, the transport equation for the macroscopic kinetic energy $\left<\rho_fk\right>=\frac{1}{2}\left<\rho_fu_iu_i\right>$ can be written as 

\begin{equation}
    \begin{split}
    \left<\rho_f\frac{d k}{dt}\right>=
    \frac{\partial \left(\phi\left<\rho_f\right>^f k_m\right)}{\partial t}
    +\frac{\partial \left(\phi \left<\rho_f\right>^f \left<u_i\right>^f k_m\right)}{\partial x_i}
    =-\phi\left<u_i\right>^f\frac{\partial \left<p\right>^f}{\partial x_i} \\
    +\phi\left<u_i\right>^f\frac{\partial \left<\tau_{ji}\right>^f}{\partial x_j} +\phi\left<\rho_f\right>^f\left<u_i\right>^f g_i+\phi\left<u_i\right>^f F_i.
    \end{split}
    \label{equ:macro_kinetic_energy}
\end{equation}
Here $\left<\rho_f k\right>$ is approximated with $\phi\left<\rho_f\right>k_m$, where $k_m=\frac{1}{2}\left<u_i\right>^{f}\left<u_i\right>^{f}$, as the dispersion part of $\rho_f\left<k\right>^f$ is neglected. In addition, it is assumed that the loss of the kinetic energy is dominated by the drag force by the porous matrix $F_i$. Thus, the viscous heat $\phi\left<\tau_{ji}\frac{\partial u_i}{\partial x_j}\right>^f$ in the macroscopic energy equation (\ref{equ:macro_energy_fluid}) is approximated with 
$\phi \left<u_i\right>^f F_i$.

The macroscopic entropy generation intensities in the flow and temperature fields can be directly calculated by averaging $s_{fD}^{gen}$ and $s_{fC}^{gen}+s_{sC}^{gen}$ in an REV. However, this requires the flow and temperature fields to be resolved in pores, which is extremely expensive. Here, we propose a macroscopic entropy transport equation, in which the macroscopic entropy generation intensity is modeled: 
Taking volume average of $\rho_f \frac{d s_f}{dt} + \rho_s\frac{d s_s}{dt}$ and considering equations (\ref{equ:EOT_f}) and (\ref{equ:EOT_s}), we can obtain  
\begin{equation}
    \left<\rho_f \frac{d s_f}{dt} + \rho_s\frac{d s_s}{dt}\right>=\left<\frac{\rho_f}{T_f}\frac{dh_f}{dt}-\frac{1}{T_f}\frac{dp}{dt}+\frac{\rho_s}{T_s}\frac{dh_s}{dt}\right>.
    \label{equ:macro_EOT}
\end{equation}
If the spatial variation of $T_f$ and $T_s$ in an REV is small, equation (\ref{equ:macro_EOT}) can be further decoupled as 
\begin{equation}
    \left<\rho_f \frac{d s_f}{dt} + \rho_s\frac{d s_s}{dt}\right>=\frac{1}{\left<T_f\right>^f}\left<\rho_f\frac{dh_f}{dt}-\frac{dp}{dt}\right>+\frac{1}{\left<T_s\right>^s}\left<\rho_s\frac{dh_s}{dt}\right>.
    \label{equ:macro_EOTB}
\end{equation}
Substituting equations (\ref{equ:macro_energy_fluid}) and (\ref{equ:macro_energy_solid}) into equation (\ref{equ:macro_EOTB}), the macroscopic transport equation of entropy can be obtained, expressed as 
\begin{equation}
\begin{split}
    \left<\rho_f \frac{d s_f}{dt} + \rho_s\frac{d s_s}{dt}\right>=\frac{\partial }{\partial x_i}\left(\frac{\Tilde{k}_f}{\left<T_f\right>^f}\frac{\partial \left<T_f\right>^f}{\partial x_i}+\frac{\Tilde{k}_s}{\left<T_s\right>^s}\frac{\partial \left<T_s\right>^s}{\partial x_i}\right) 
    +\alpha A_v\frac{\left(\left<T_s\right>^s-\left<T_f\right>^f\right)^2}{\left<T_s\right>^s \left<T_f\right>^f}
    \\
    +\frac{\Tilde{k}_f}{\left<T_f\right>^{f2}}\frac{\partial \left<T_f\right>^f}{\partial x_i}\frac{\partial \left<T_f\right>^f}{\partial x_i}
    +\frac{\Tilde{k}_s}{\left<T_s\right>^{s2}}\frac{\partial \left<T_s\right>^s}{\partial x_i}\frac{\partial \left<T_s\right>^s}{\partial x_i}
    +\frac{\phi \left<u_i\right>^f F_i}{\left<T_f\right>}.
   \end{split}
    \label{equ:macro_entropy}
\end{equation}

Equation (\ref{equ:macro_entropy}) indicates that entropy is generated in an isolated system due to heat transfer between the fluid and solid phases, heat transfer within the fluid and solid phases, and pressure drop (or dissipation) in the flow field. The corresponding entropy generation intensities $\dot{s}^{mac}_{hsf}$, $\dot{s}^{mac}_{hts}$, $\dot{s}^{mac}_{htf}$ and $\dot{s}^{mac}_{pre}$ are 
\begin{equation}
\dot{s}^{mac}_{hsf}= \alpha A_v \frac{\left(\left<T_s\right>^s-
        \left<T_f\right>^f\right)^2}{\left<T_s\right>^s \left<T_f\right>^f}, 
\end{equation}

\begin{equation}
\dot{s}^{mac}_{htf}=\frac{\Tilde{k}_f}{\left<T_f\right>^{f2}}\frac{\partial \left<T_f\right>^f}{\partial x_i}\frac{\partial \left<T_f\right>^f}{\partial x_i},     
\end{equation}

\begin{equation}
\dot{s}^{mac}_{hts}=\frac{\Tilde{k}_s}{\left<T_s\right>^{s2}}\frac{\partial \left<T_s\right>^s}{\partial x_i}\frac{\partial \left<T_s\right>^s}{\partial x_i}, 
\end{equation}
and 
\begin{equation}
\dot{s}^{mac}_{pre}=\frac{\phi \left<u_i\right>^f F_i}{\left<T_f\right>}, 
\end{equation}
respectively. 

Similarly, taking volume average of $\rho_f\frac{db_f}{dt}+\rho_s\frac{dbs}{dt}$ and considering equations (\ref{equ:EOT_exergy_f}) and (\ref{equ:EOT_exergy_s}), there is
\begin{equation}
    \left<\rho_f \frac{d b_f}{dt} + \rho_s\frac{d b_s}{dt}\right>=\left<\rho_f\frac{dh_f}{dt}+\rho_s\frac{dh_s}{dt}-T_0\left(\rho_f\frac{ds_f}{dt}+\rho_s\frac{ds_s}{dt}\right)+\rho_f\frac{dk}{dt}\right>.
    \label{equ:macro_EOT_exergy}
\end{equation}
Substituting equations (\ref{equ:macro_energy_fluid}), (\ref{equ:macro_energy_solid}), 
(\ref{equ:macro_kinetic_energy}) and
(\ref{equ:macro_entropy}) into equation (\ref{equ:macro_EOT_exergy}), the following macroscopic transport equation of exergy can be obtained, expressed as, 

\begin{equation}
\begin{split}
    \left<\rho_f \frac{d b_f}{dt} + \rho_s\frac{d b_s}{dt}\right>=\phi\frac{\partial \left<p\right>^f}{\partial t}
    +\frac{\partial }{\partial x_i}\left(\Tilde{k}_f\left(1-\frac{T_0}{\left<T_f\right>^f}\right)\frac{\partial \left<T_f\right>^f}{\partial x_i}\right)
    \\
    +\frac{\partial }{\partial x_i}\left(\Tilde{k}_s\left(1-\frac{T_0}{\left<T_s\right>^s}\right)\frac{\partial \left<T_s\right>^s}{\partial x_i}\right) 
    -    T_0 \left(\dot{s}^{mac}_{hsf}+\dot{s}^{mac}_{htf}+\dot{s}^{mac}_{hts}+\dot{s}^{mac}_{pre}\right).
        \end{split}
    \label{equ:macro_exergy}
\end{equation}

Equation (\ref{equ:macro_exergy}) shows that, the exergy is destructed due to the entropy generation, the corresponding exergy loss coefficients are termed as thermal loss coefficient
    \begin{equation}
    \zeta^b_{hsf}=\frac{1}{B_{chg}}\int\limits_{cyc}\int\limits_{vol}\dot{l}_{hsf}dVdt=\frac{1}{B_{chg}}\int\limits_{cyc}\int\limits_{vol}T_0 \dot{s}^{mac}_{hsf}dVdt,   
    \end{equation}
fluid conduction loss coefficient
\begin{equation}
\zeta^b_{htf}=\frac{1}{B_{chg}}\int\limits_{cyc}\int\limits_{vol}\dot{l}_{htf}dV=\frac{1}{B_{chg}}\int\limits_{cyc}\int\limits_{vol}T_0 \dot{s}^{mac}_{htf}dVdt,
\end{equation}
solid conduction loss coefficient
\begin{equation}
\zeta^b_{hts}=\frac{1}{B_{chg}}\int\limits_{cyc}\int\limits_{vol}\dot{l}_{hts}dV=\frac{1}{B_{chg}}\int\limits_{cyc}\int\limits_{vol}T_0 \dot{s}^{mac}_{hts}dVdt,    
\end{equation}
and pressure loss coefficient
\begin{equation}
\zeta^b_{pre}=\frac{1}{B_{chg}}\int\limits_{cyc}\int\limits_{vol}\dot{l}_{pre}dV=\frac{1}{B_{chg}}\int\limits_{cyc}\int\limits_{vol}T_0 \dot{s}^{mac}_{pre}dVdt.    
\end{equation}
where $B_{chg}$ is the total amount of exergy entering the tank from the inlet (top tank surface). $\dot{l}_{hsf}$, $\dot{l}_{htf}$, $\dot{l}_{hts}$ and $\dot{l}_{pre}$ are the (local) exergy loss intensities due to the heat transfer between solid and fluid, heat transfer in the fluid, heat transfer in the solid and pressure loss, respectively. Integrating them in the tank volume yields the exergy loss rates $\dot{L}_{hsf}$, $\dot{L}_{htf}$, $\dot{L}_{hts}$ and $\dot{L}_{pre}$. Besides these losses that occur inside the the porous medium, the exergy is also lost at the boundary surfaces. The corresponding loss coefficients are the exit loss coefficient (due to the outflow of the HTF at the bottom tank)
\begin{equation}
    \zeta^b_{bot}=\frac{1}{B_{chg}}\int\limits_{cyc}\dot{L}^b_{bot}dt=\frac{1}{B_{chg}}\int\limits_{cyc}\int\limits_{bot}{\phi \left<\rho_f\right>^f \left<u_i\right>^f \left<b_f\right>^f}n_idAdt, 
\end{equation}
and the heat leakage loss coefficient (due to the heat release at the wall surfaces) 
\begin{equation}
\begin{split}
\zeta^b_{wal}=\frac{1}{B_{chg}}\int\limits_{cyc}\dot{L}^b_{wal}dt
=\frac{1}{B_{chg}}
\int\limits_{cyc}\int\limits_{wal}
    \left({\Tilde{k}_f\left(1-\frac{T_0}{\left<T_s\right>^f}\right)\frac{\partial \left<T_f\right>^f}{\partial x_i}}
   +{\Tilde{k}_s\left(1-\frac{T_0}{\left<T_s\right>^s}\right)\frac{\partial \left<T_s\right>^s}
    {\partial x_i}}\right)
    n_idAdt.    
\end{split}
\end{equation}
The total exergy loss coefficient, denoted by
\begin{equation}
    \zeta^b_{tot}=\zeta^b_{hsf}+\zeta^b_{htf}+\zeta^b_{hts}+\zeta^b_{pre}+\zeta^b_{bot}+\zeta^b_{wal}
\end{equation}
is the sum of the exergy loss coefficients due to to the aforementioned mechanisms. 
These loss coefficients are termed to be consistent with the definitions by White, et al. (2016) \cite{whiteAnalysisOptimisationPackedbed2016}, while they can be used to perform the SLA of multi-dimensional CFD results.

\subsection{Numerical methods}
This study solves the macroscopic equations (\ref{equ::macro_mass})-(\ref{equ:macro_energy_solid}). Pore-scale geometry of the packed-bed is modeled, not directly resolved. A finite volume method is used for the simulation. The solver is developed based on the conjugate heat transfer solver chtMultiRegionFoam from the open-source code package OpenFOAM 22.12. An Euler implicit scheme is used for the time discretization. A second-order central difference scheme is used for the convective terms in the momentum and energy equations of the fluid phase. A second order central difference scheme is used for the diffusion terms in both the fluid and solid phases. The solid and fluid regions completely overlap. The temperature of each phase is interpolated into the other phase to calculate the heat source term due to the heat transfer rate between the two phases. 
 
\section{Description of test cases and validation}
\begin{figure}
    \centerline{\includegraphics[width=100mm]{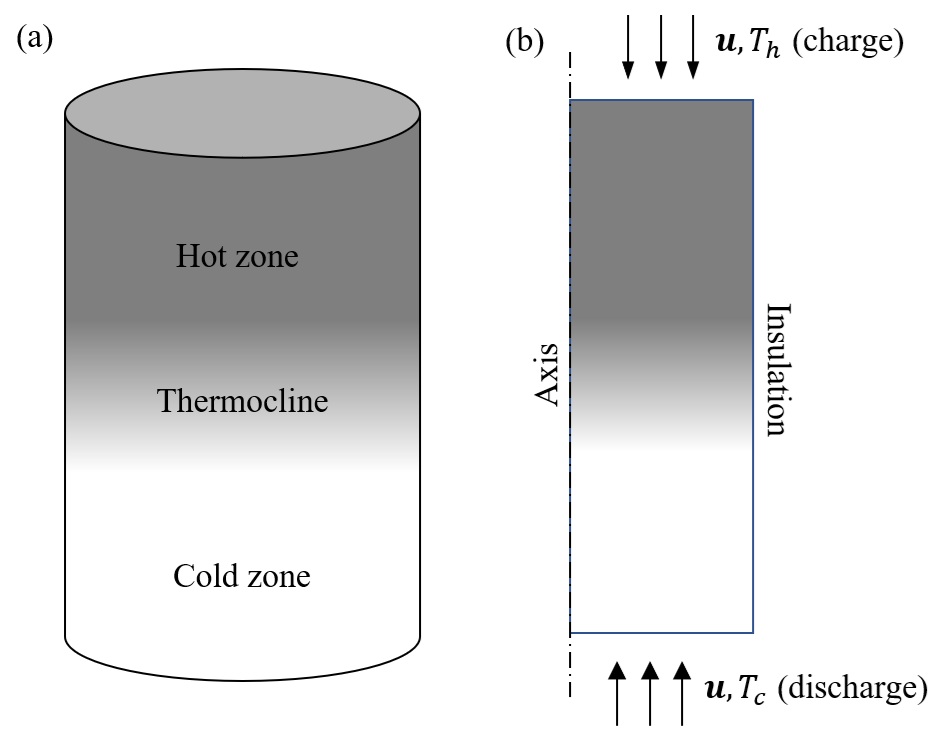}}
    \caption{Schematic illustration of a packed-bed thermal store (a) and the corresponding two-dimensional axisymmetric computational domain.}
    \label{fig:domain}
\end{figure}

To demonstrate the application of the SLA to system optimization, we have simulated the charge and discharge process of a laboratory-scale packed-bed SHS system. The experiment was performed in the PROMES-CNRS laboratory and reported in \cite{hoffmannThermoclineThermalEnergy2016}. In the experiment, the storage tank, which is filled with quartzite rocks (TESM), is heated and cooled with rapeseed oil. The oil acts as the heat transfer fluid (HTF) and flows through the tank. A schematic illustration of the studied SHS system is shown in figure \ref{fig:domain}. The computational domain is assumed to be axi-symmetric to reduce the computational cost. The main parameters used in the experiments are shown in table \ref{tab:Promes}. The properties of the HTF and the TESM are shown in table \ref{tab:properties} .

\begin{table}
  \begin{center}
\def~{\hphantom{0}} 
  \begin{tabular}{llllllll}
  \hline
      Energy      &&  $8.3 \, \mathrm{kWh_T}$             \\
      Discharge time $t_{dis}$   &&  $3 \, \mathrm{h}$    \\
      $^{*}$Charge time $t_{chg}$  &&  $3 \, \mathrm{h}$    \\
      Tank height $H$  &&  $1.8 \, \mathrm{m}$ \\
      Tank diameter $D$  && $0.4 \, \mathrm{m}$ \\
      Tank volume $V$ && $0.23 \, \mathrm{m^3}$ \\
      Porosity $\phi$ && $0.41$ \\
      Particle diameter $d_p$ && $40 \, \mathrm{mm}$ \\
      HTF mass flow rate $\dot{m}$ && $0.019 \, \mathrm{kg\,m^{-1}}$ \\
      HTF high temperature $T_H$ && $483 \, \mathrm{K}$ \\
      HTF low temperature $T_L$ && $433 \, \mathrm{K}$ \\
      \hline
      \\
   \end{tabular}
   
  \caption{Main parameters used in the experiments by the PROMES-CNRS laboratory. The parameter with $^*$ is tentatively given in this study.}
  \label{tab:Promes}
  \end{center}
\end{table}

\begin{table}
  \begin{center}
\def~{\hphantom{0}}
  \begin{tabular}{llllllll}
   \hline
       HTF density $\rho_f$      &&  $804$ && $\mathrm{kg\,m^{-3}}$             \\
      TESM density $\rho_s$      &&  $2500$ && $\mathrm{kg\,m^{-3}}$             \\      
      HTF heat capacity $C_{pf}$ &&  $2086-0.84T_f$ &&  $\mathrm{J \, kg^{-1}\,K^{-1}}$             \\
      TESM heat capacity $C_{ps}$ &&  $830$ &&  $\mathrm{J \, kg^{-1}\,K^{-1}}$             \\
      HTF heat conductivity $k_f$ &&  $0.0981-2.4\times10^{-4}T_f$ && $\mathrm{W\,m^{-1}\,K^{-1}}$     \\
      TESM heat conductivity $k_s$ &&  $5.69$ && $\mathrm{W\,m^{-1}\,K^{-1}}$     \\
      HTF dynamic viscosity $\mu_f$ && $0.02184-3.9\times0^{-5}T_f$ && $\mathrm{Pa\,s}$ \\
  \hline
  \\ 
   \end{tabular}
    \\
  \caption{Thermal properties of the HTF and TESM.}
  \label{tab:properties}
  \end{center}
\end{table}

The tank is insulated with $\delta_w=20 \, \mathrm{cm}$ of rock wool covered with aluminum. The insulation effect is modeled at the the wall surface. The corresponding boundary conditions for the fluid and solid phases are
\begin{equation}
    \tilde{k}_f \mathbf{n} \cdot \mathbf{\nabla}T_f=k_w\frac{T_o-T_{wf}}{\delta_w}=\phi \alpha_a\left(T_a-T_o\right), 
    \label{wbc_fluid}
\end{equation}
and
\begin{equation}
    \tilde{k}_s \mathbf{n} \cdot \mathbf{\nabla}T_s=k_w\frac{T_o-T_{ws}}{\delta_w}=\left(1-\phi\right) \alpha_a\left(T_a-T_o\right).
    \label{wbc_solid}
\end{equation}
Here $T_{wf}$ and $T_{ws}$ are the inner wall surface temperatures of the fluid and solid phases, respectively. It is assumed that their difference is small. $T_o$ is the external surface temperature of the insulation. $T_a$ is the ambient temperature, which is assumed to be $300\,K$. The heat transfer coefficient from the external wall to the ambient $h_a$ is estimated to be $0.63\,\mathrm{Wm^{-1}K^{-1}}$ according to \cite{hoffmannThermoclineThermalEnergy2016}. 

\begin{figure}
    \centerline{\includegraphics[width=100mm]{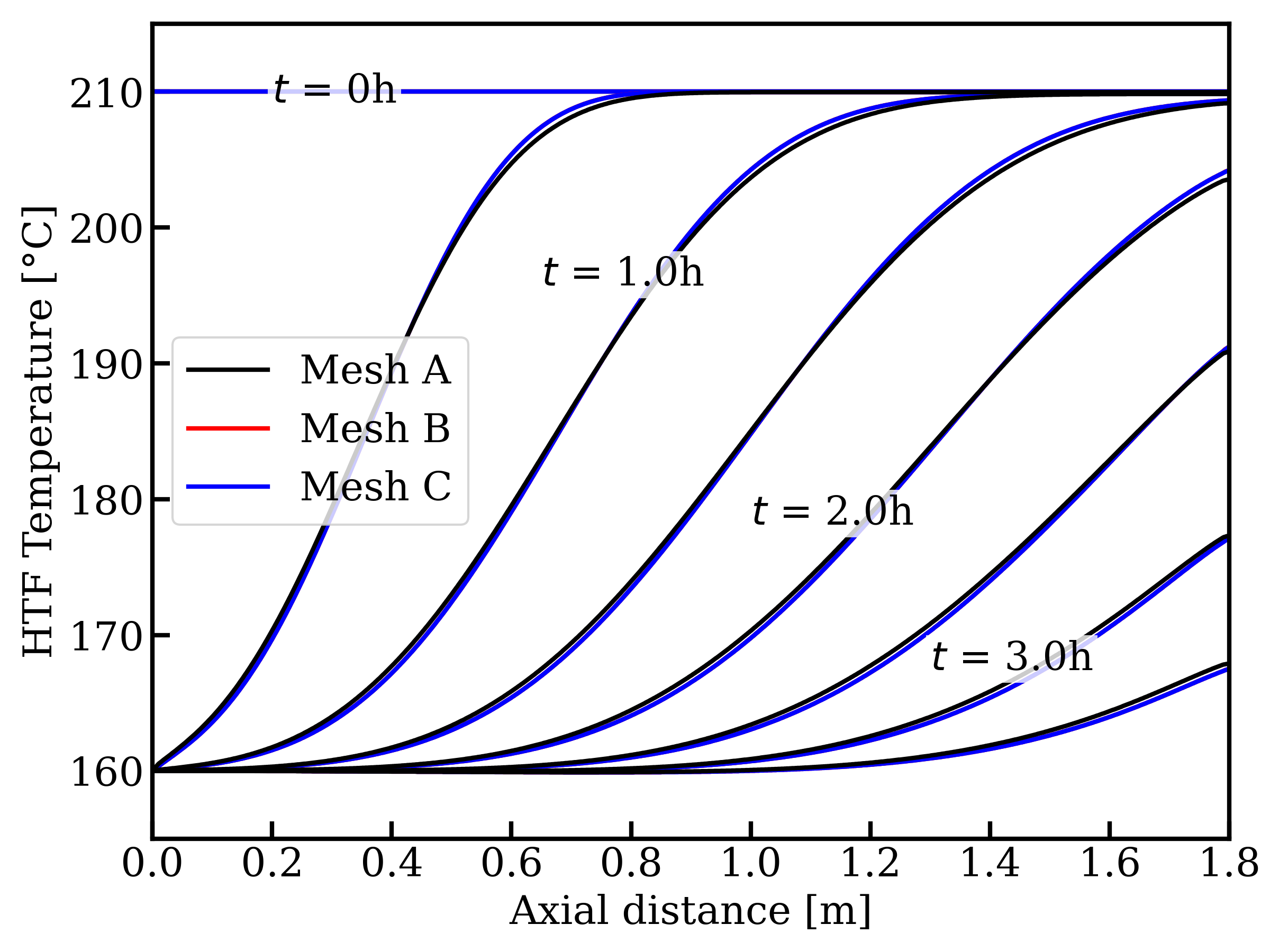}}
    \caption{Time evolution of the HTF temperature at the center line. The numerical results of mesh A($20\times180$), B($30\times270$) and C($40\times360$) are compared.}
    \label{fig:verification}
\end{figure}

A slip boundary condition is applied at the tank boundary walls. The viscous layer near the wall surfaces is neglected since its effect is limited to a very thin boundary layer close to the wall \cite{Nield2017}. The study uses three mesh resolutions to examine their effect on the numerical simulations, they give 100 (mesh A), 150 (mesh B) and 200 (mesh C) mesh cells per meter. Figure \ref{fig:verification} shows the time evolution of the HTF temperature at the tank centerline calculated using the three mesh resolutions. It can be seen that the numerical results from the three mesh resolutions are almost identical. The highest mesh resolution (mesh C) is adopted in this study to ensure the mesh-independence of the numerical results. 

\begin{figure}
    \centerline{\includegraphics[width=100mm]{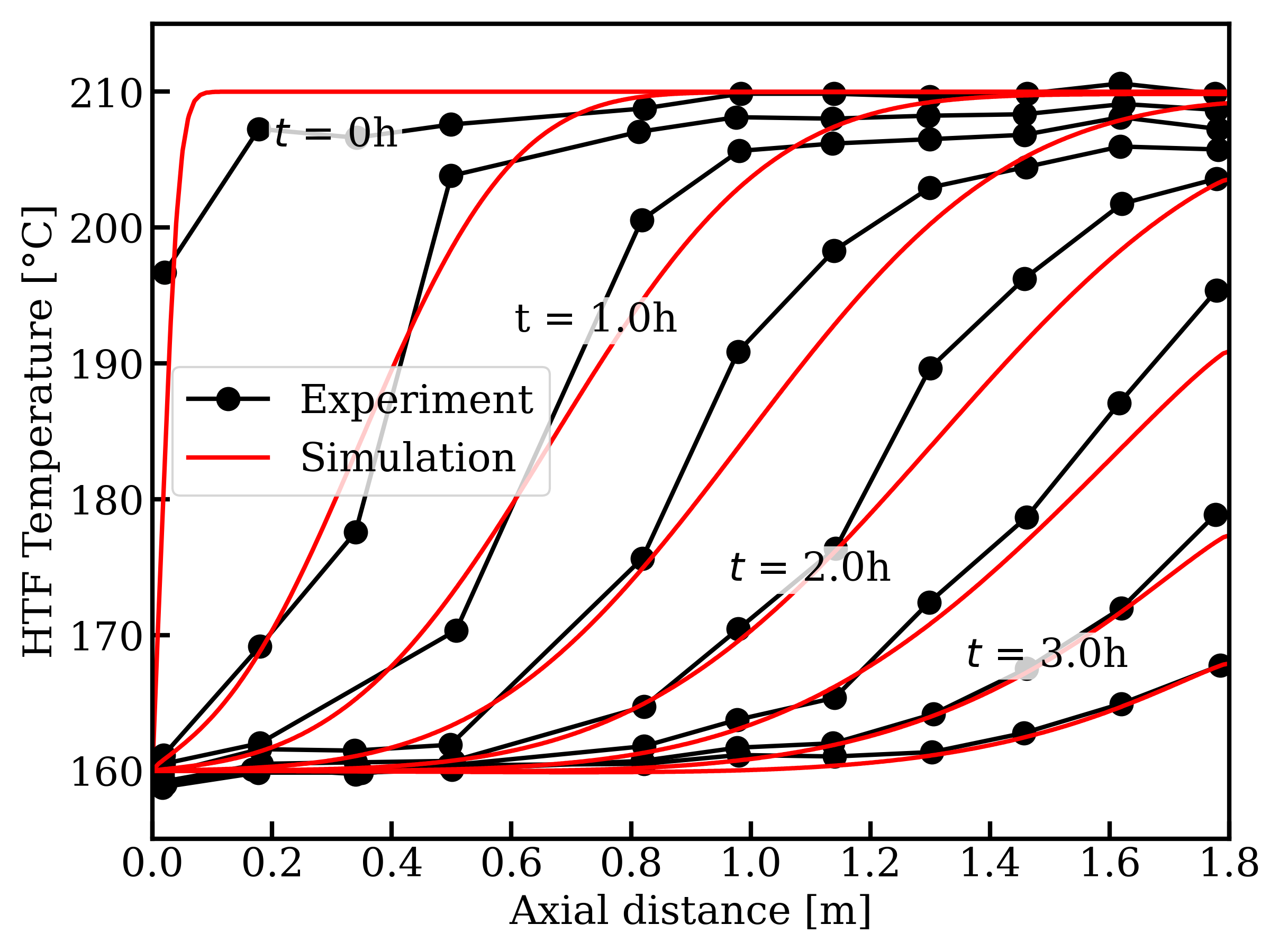}}
    \caption{Time evolution of the HTF temperature at the center line. The numerical results are compared with the experimental results in \cite{hoffmannThermoclineThermalEnergy2016}.}
    \label{fig:validation}
\end{figure}

The numerical results are then compared with the experimental data in \cite{hoffmannThermoclineThermalEnergy2016}, see figure \ref{fig:validation}. There is a slight discrepancy between the two results, as the experiment has not been fully represented in the simulation. For instance, the heat release at the wall is modeled using zero-dimensional equations (\ref{wbc_fluid}) and (\ref{wbc_solid}), while the heat conduction within the insulating walls is neglected. However, it is expected that this difference will not affect our analysis.

\section{Results and discussion}

\begin{figure}
    \centerline{\includegraphics[width=125mm]{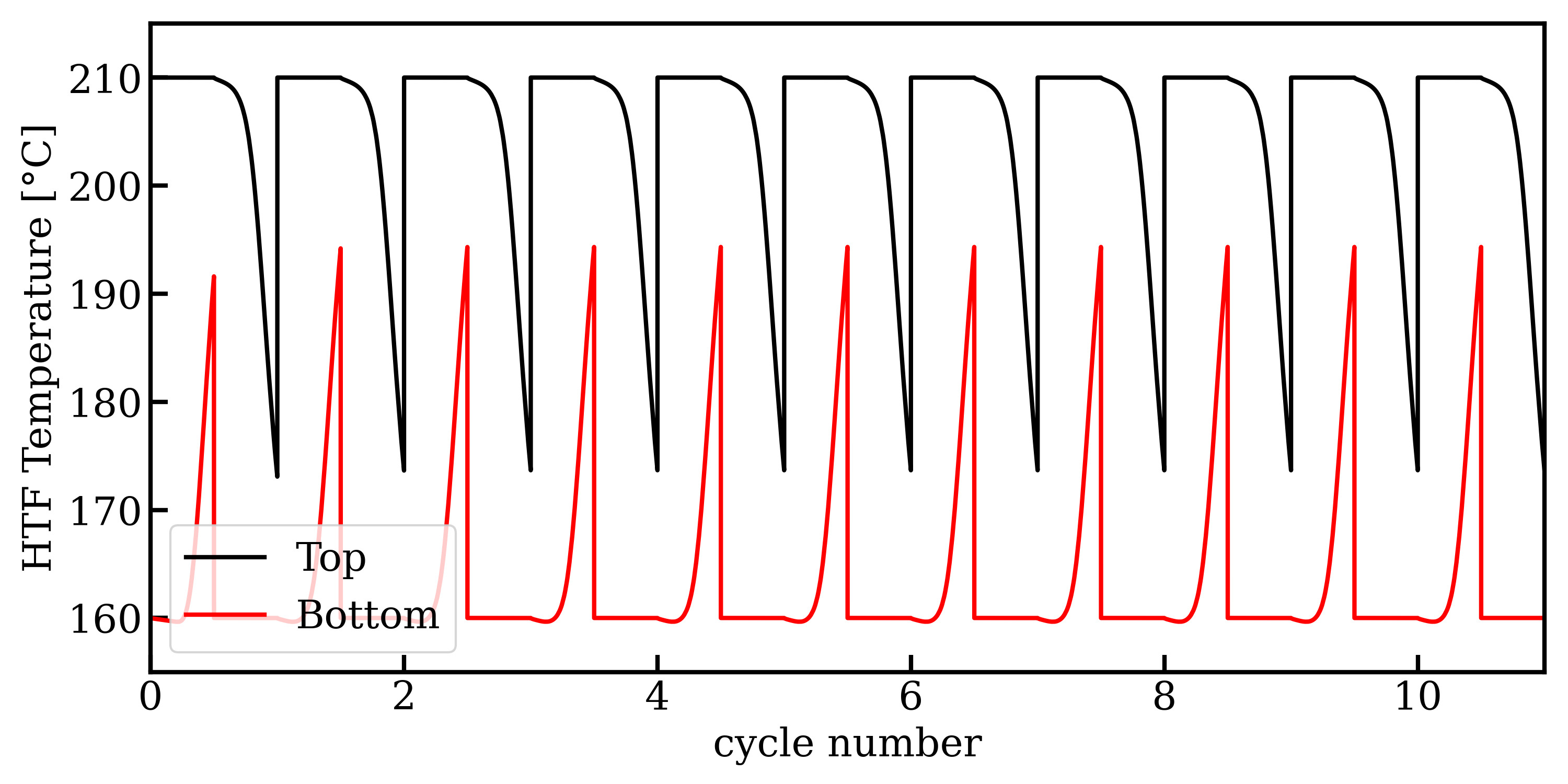}}
    \caption{Time evolution of the averaged HTF temperature at the top and bottom of the storage tank.}
    \label{fig:charge_discharge}
\end{figure}

\begin{figure}
    \centerline{\includegraphics[width=100mm]{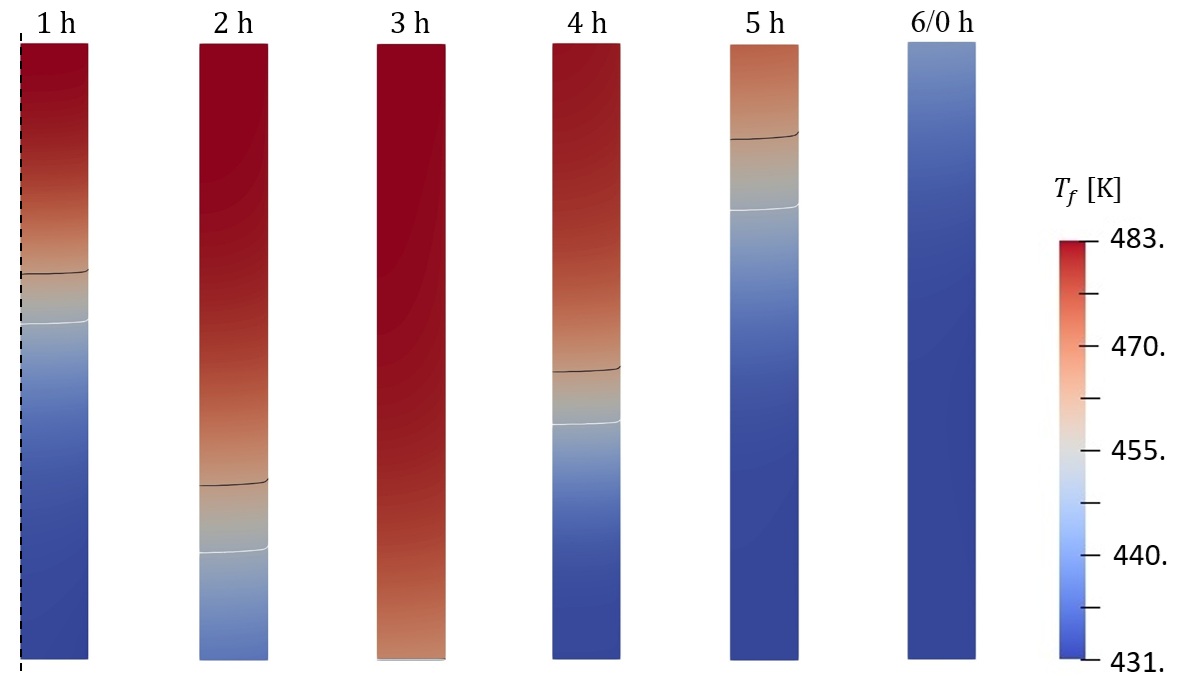}}
    \caption{Fluid temperature $T_f$ fields during charge ($0\mathrm{h}$-$3\mathrm{h}$) and discharge ($3\mathrm{h}$-$6\mathrm{h}$). The iso-surfaces of $T=465 \mathrm{K}$ (black lines) and $T=455 \mathrm{K}$ (white lines) are shown to indicate the location of the thermocline. The axis of the tank is indicated at $T=1\, h$}
    \label{fig:T_charge_discharge}
\end{figure}

Hoffmann, et al. (2016) only reported the experimental data for the three-hour discharge process \cite{hoffmannThermoclineThermalEnergy2016}. We assume that the charging process also takes place over the course of three hours. Similar to \cite{McTigue2018}, we have neglected the heat released during the storing time between charge and discharge. Ten charge-discharge cycles are calculated, see figure \ref{fig:charge_discharge}. A fully developed cycle, which is also called a steady-state cycle in \cite{McTigue2018}, is obtained by the third cycle. The flow and temperature fields change periodically once the charge-discharge process is fully developed. We use the results of one fully developed cycle in our analysis. Figure \ref{fig:T_charge_discharge} shows the fluid temperature fields in a fully-developed charge-discharge cycle. For simplicity, the starting time of this cycle is set to $t=0\,\mathrm{h}$, whose results are the same as those at the end of the cycle ($t=6\,\mathrm{h}$). The thermocline indicated by the region between $T_f=455\,\mathrm{K}$ and $T_f=465\,\mathrm{K}$ can be clearly seen during the charge ($t=1\,h$ and $2\,\mathrm{h}$) and discharge ($t=4\,\mathrm{h}$ and $5\,\mathrm{h}$). The thermocline in the solid phase is very similar to the thermocline in the fluid phase; therefore, they are assumed to be the same. 

\subsection{Energy and exergy loss coefficients}
\begin{figure}
    \centerline{\includegraphics[width=80mm]{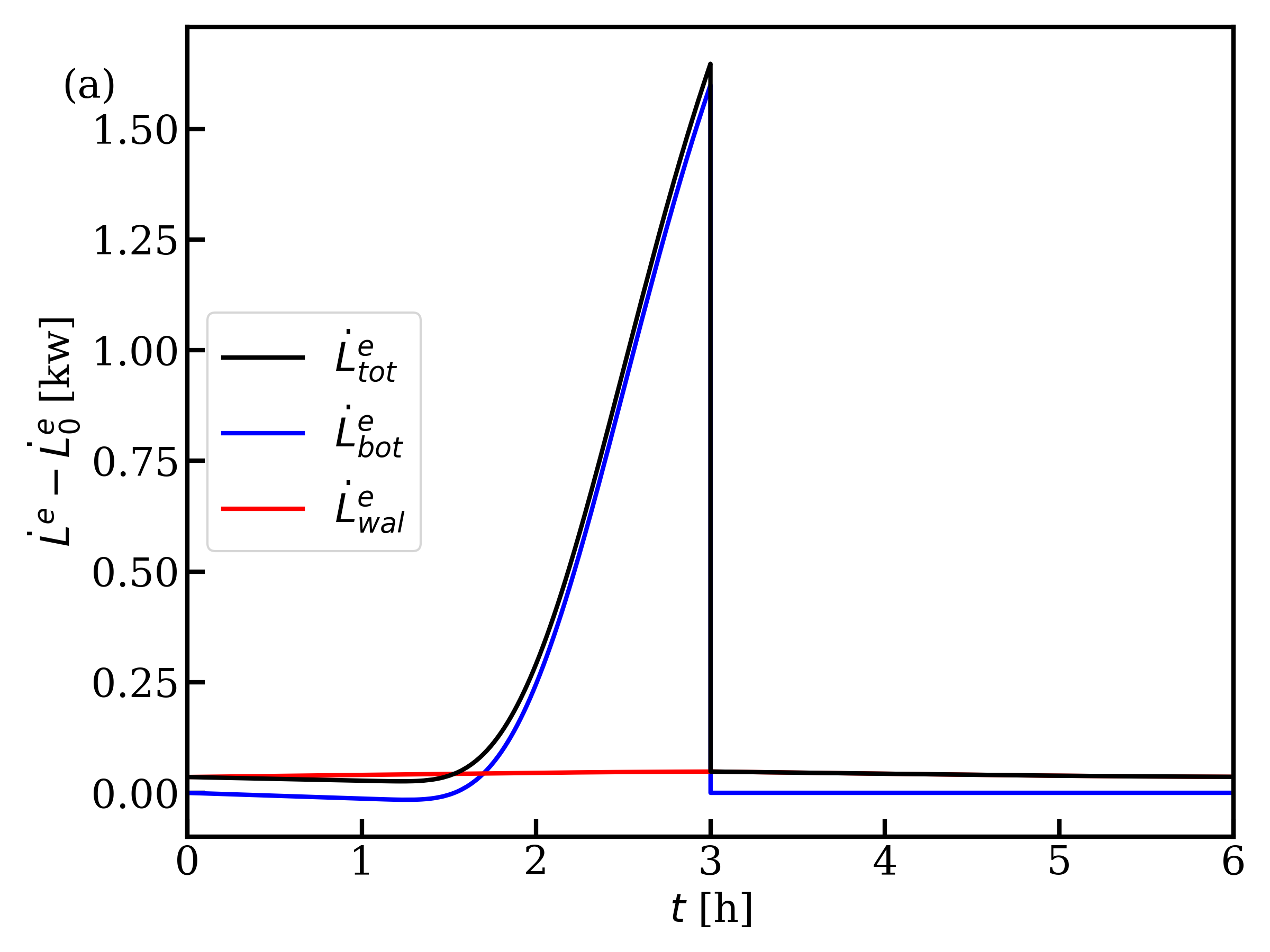}
                \includegraphics[width=80mm]{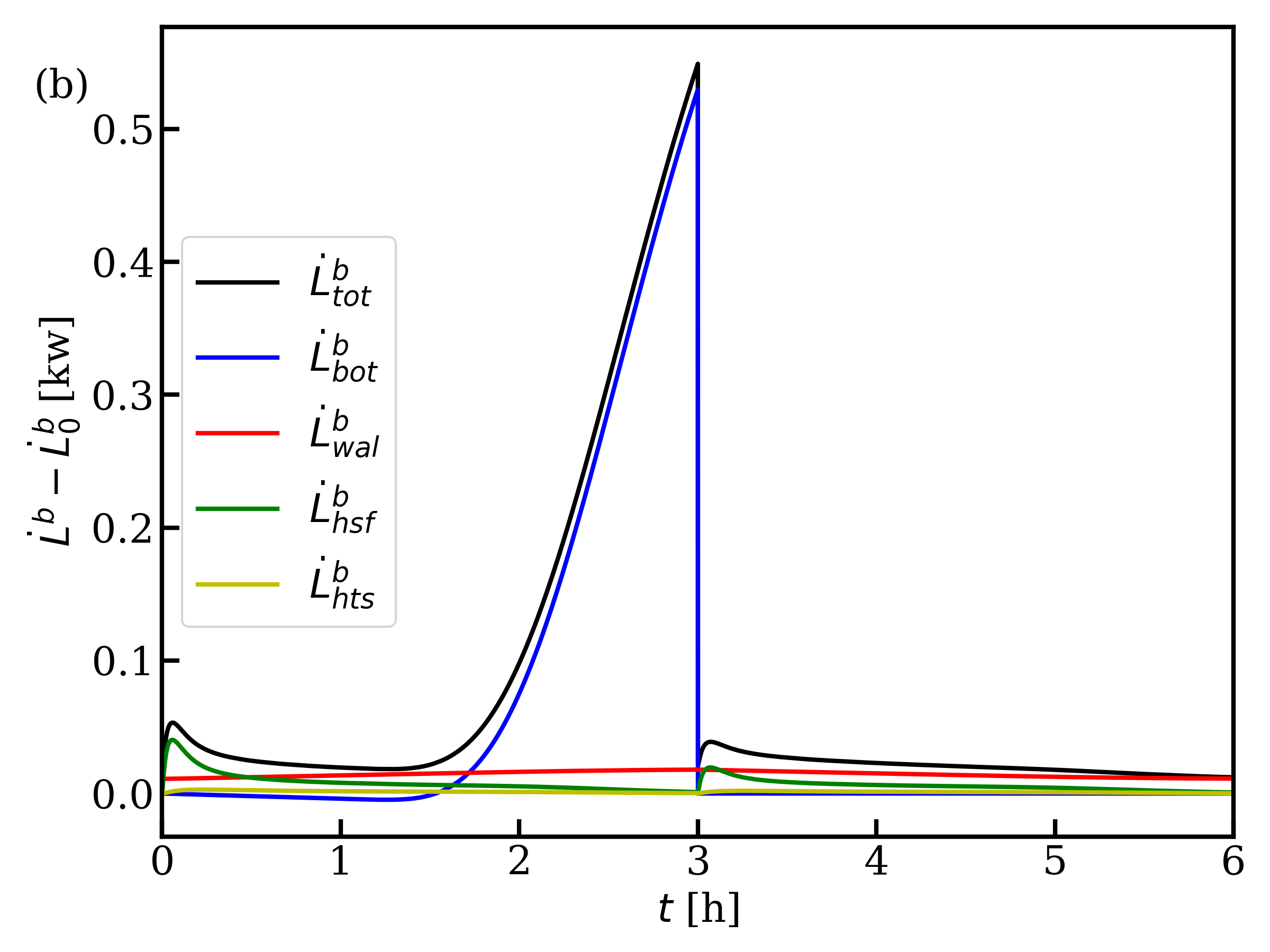}}
   \caption{Energy (a) and exergy (b) loss rates due to different mechanisms in a fully-developed charge-discharge cycle.}
    \label{fig:loss_Promes}
\end{figure}

Figure \ref{fig:loss_Promes} shows the (thermal) energy and exergy loss rates during a fully developed cycle. Energy loss is primarily due to the heat released from the tank bottom ($\dot{L}^e_{bot}$), which occurs during charge. Heat leakage loss, due to the heat release at the tank wall surface ($\dot{L}^e_{wal}$), is significantly smaller. However, as stated in \cite{whiteAnalysisOptimisationPackedbed2016}, exit loss cannot always be considered as a real loss since the HTF may be transferred to other components in the cycle.

\begin{figure}
    \centerline{\includegraphics[width=100mm]{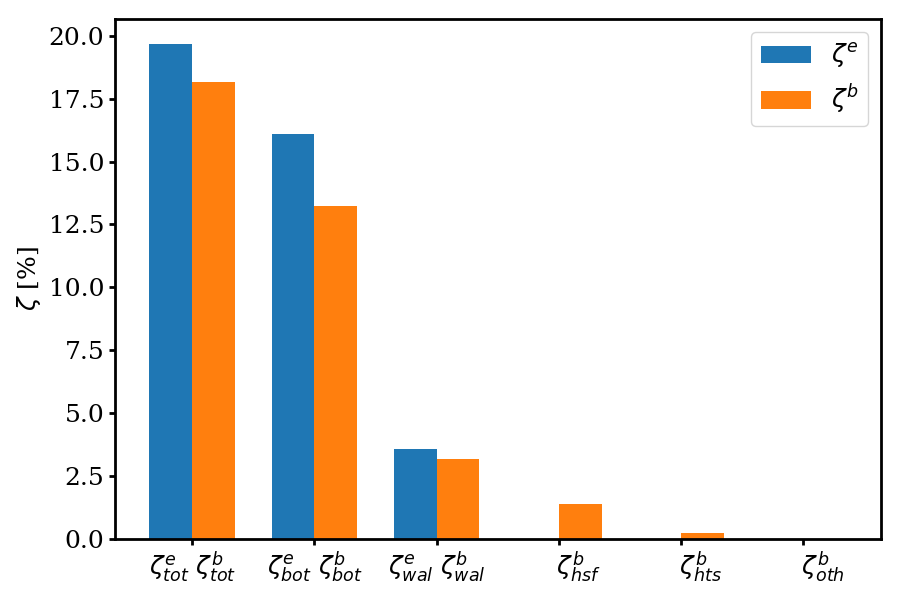}}
    \caption{Total energy and exergy loss coefficients and those due to different mechanisms.}
    \label{fig:loss_Promes_comparison}
\end{figure}

In addition to the exit and heat leakage losses, the SLA suggests that the exergy is also destructed due to the heat transfer between the fluid and solid (thermal loss $\dot{L}_{hsf}^b$) and the heat conduction in the solid (solid conduction loss $\dot{L}_{hts}^b$), see figure \ref{fig:loss_Promes}b. The other losses, including the fluid conduction loss $\dot{L}_{htf}^b$ and the pressure loss $\dot{L}_{pre}^b$, are negligibly small. A comprehensive comparison of the energy and exergy loss coefficients is shown in figure \ref{fig:loss_Promes_comparison}, where the energy loss coefficients are calculated as $\zeta^e_{wal}=\frac{1}{Q_{chg}}\int\limits_{cyc}\dot{L}^e_{wal}dt$ and $\zeta^e_{bot}=\frac{1}{Q_{chg}}\int\limits_{cyc}\dot{L}^e_{bot}dt$. 

\begin{figure}
    \centerline{\includegraphics[width=100mm]{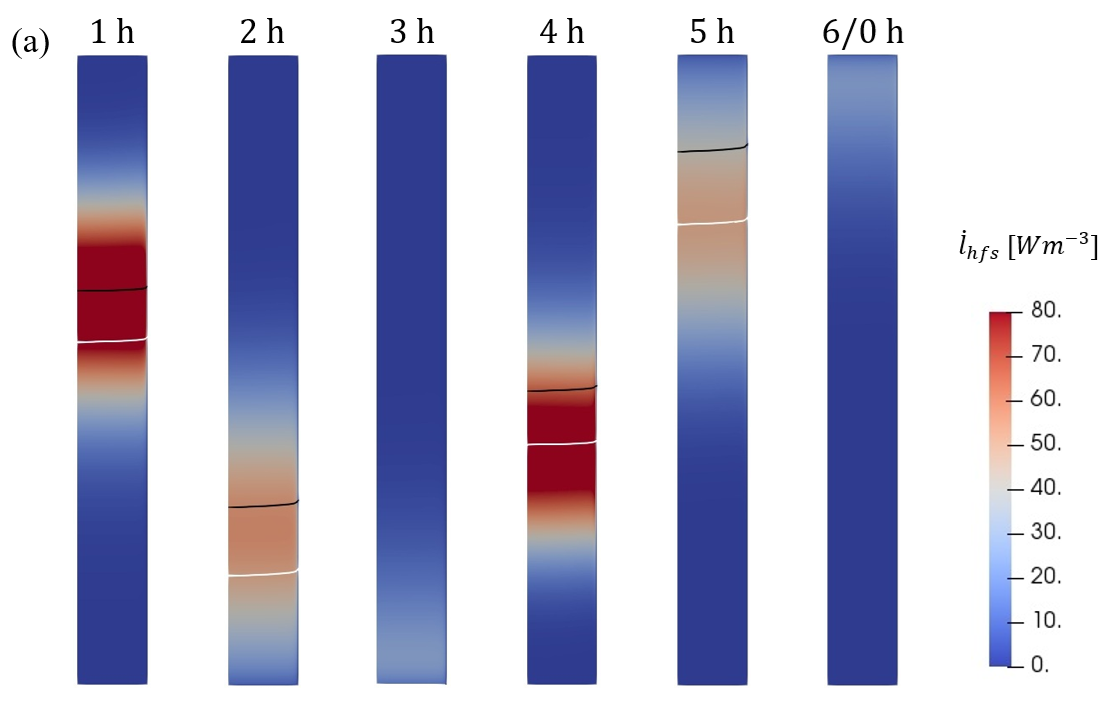}}
    \centerline{\includegraphics[width=100mm]{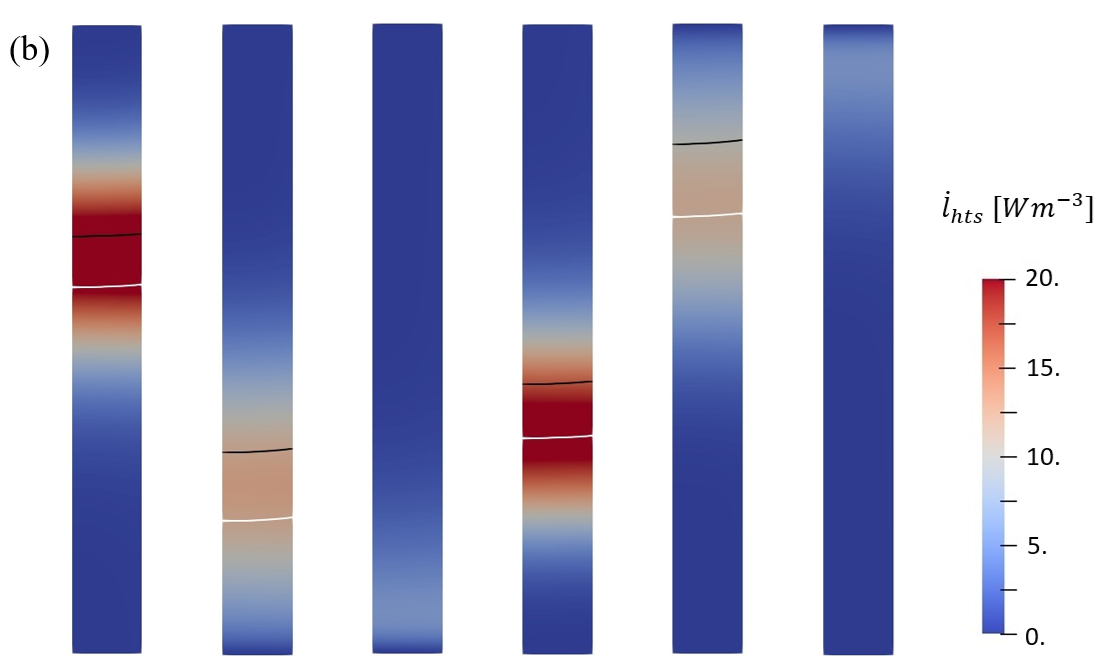}}
    \caption{Thermal loss intensity $\dot{l}_{hsf}$ (a) and solid conduction loss intensity $\dot{l}_{hts}$ during the charge ($0\mathrm{h}$-$3\mathrm{h}$) and discharge ($3\mathrm{h}$-$6\mathrm{h}$). The iso-surfaces of $T=465 \mathrm{K}$ (black lines) and $T=455 \mathrm{K}$ (white lines) are shown to indicate the location of the thermocline.}
    \label{fig:loss_Promes_2d}
\end{figure}

Based on the numerical results, we can determine where the losses occur and how strong they are. It is already known that the exit loss occurs at the bottom of the tank and the heat leakage loss occurs at the wall surface. The other losses occur within the tank. Figure \ref{fig:loss_Promes_2d} shows the instantaneous exergy loss intensities, $\dot{b}_{hsf}^b$ and $\dot{b}_{hts}^b$, at typical time instants during the charge ($t=1, 2, 3 \mathrm{h}$) and discharge ($t=4, 5, 6 \mathrm{h}$) processes. As can be seen in the figure, the thermal and the solid conduction losses occur very close to the thermocline indicated by the region between the iso-surfaces of $T=455\,K$ and $T=465\,K$. Both $\dot{b}_{hsf}^b$ and $\dot{b}_{hts}^b$ are larger near the beginning of the charge or discharge processes, suggesting stronger heat transfer during these periods. The thermocline moves slightly faster than $\dot{b}_{hsf}^b$ and $\dot{b}_{hts}^b$. This occurs because the heat transfer between the fluid and solid, as well as the heat conduction within the solid, only happen once the surrounding fluid has reached a high temperature. Additionally, it can be also seen that the thermocline is slightly inclined to the wall surface, which is due to the heat release at the wall.

\subsection{Effect of tank aspect ratio}
\begin{figure}
    \centerline{\includegraphics[width=80mm]{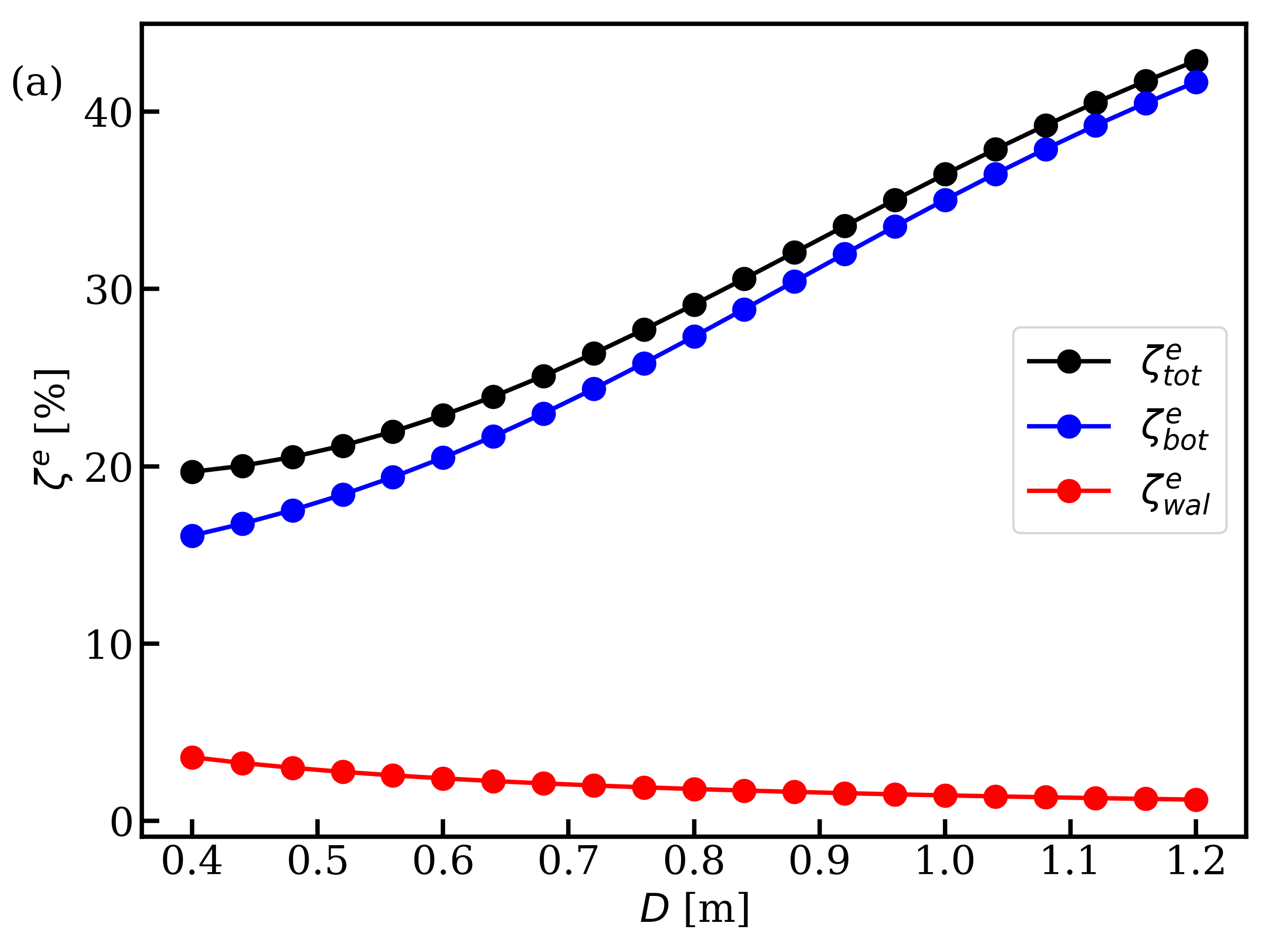}
                \includegraphics[width=80mm]{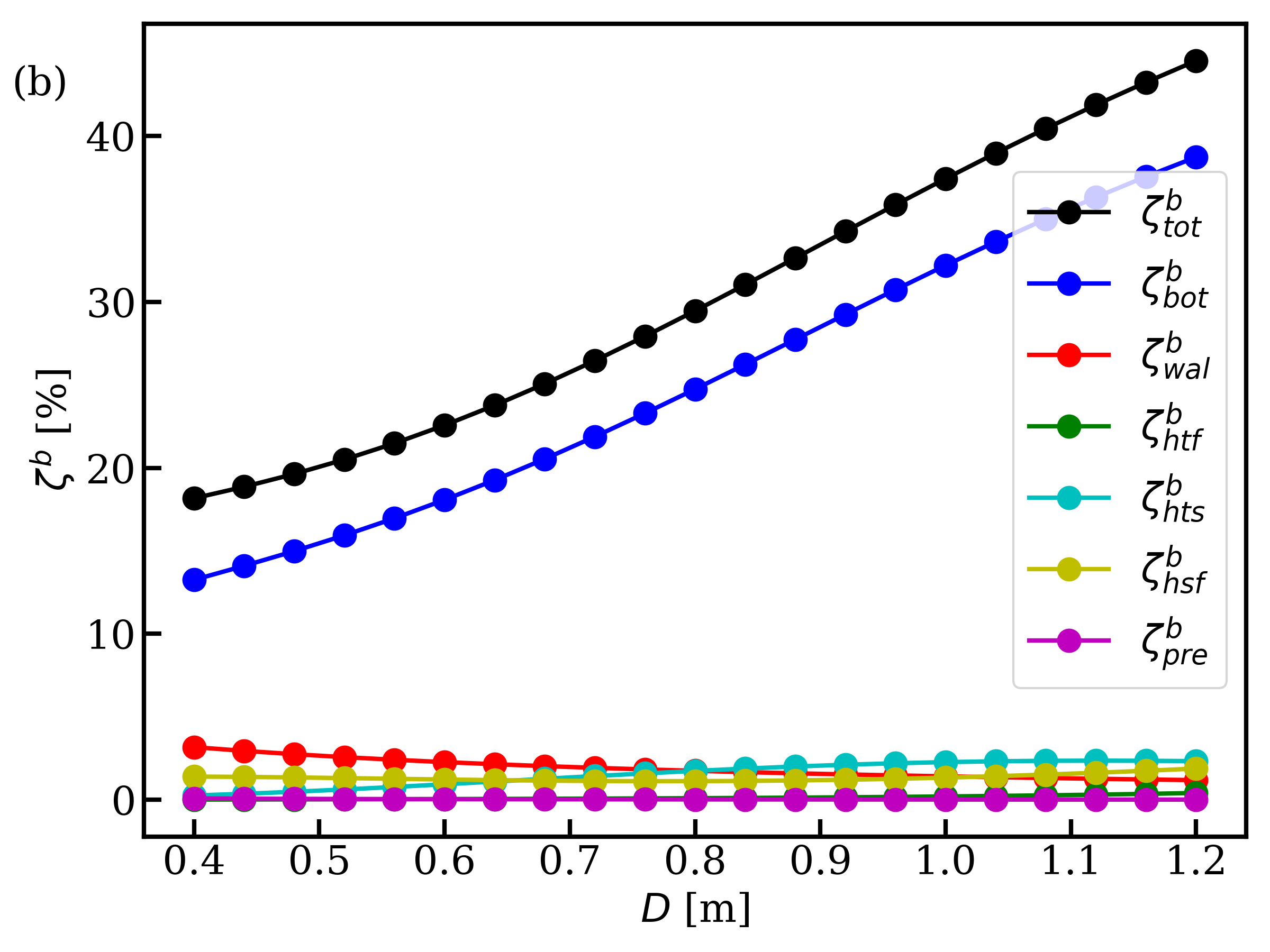}}
    \caption{Effect of the tank diameter $D$ on the energy (a) and exergy (b) loss coefficients due to different mechanisms. The tank volume is kept as a constant ($0.23 \, \mathrm{m^3}$). The exit loss is taken into account in the total loss.}
    \label{fig:loss_HD_exit}
\end{figure}

To reduce the total (energy or exergy) loss, we have kept the tank volume at the original value $V=0.23\,\mathrm{m^3}$, then studied the effect of the tank aspect ratio $D/H$ on the loss coefficients. The studied tank diameter $D$ ranges from $0.4\,\mathrm{m}$ to $1.2\, \mathrm{m}$. The corresponding tank height ranges from $1.8\,\mathrm{m}$ to $0.2\,\mathrm{m}$. Figure \ref{fig:loss_HD_exit} shows that, the energy exit loss coefficient $\zeta_{bot}^{e}$ increases as $D$ increases. Meanwhile, the heat leakage loss coefficient $\zeta_{wal}^{e}$ decreases as the wall surface area decreases. The energy analysis favors a low tank aspect ratio $D/H$ for the parameters under consideration. It is expected that there is a minimum value of $\zeta^e_{tot}$ at a vary small tank diameter, which is not considered in this study. Note that this study does not consider the storage time between the end of charging and the start of discharging. If this is taken into account, the heat leakage loss could be more significant.

\begin{figure}
    \centerline{\includegraphics[width=100mm]{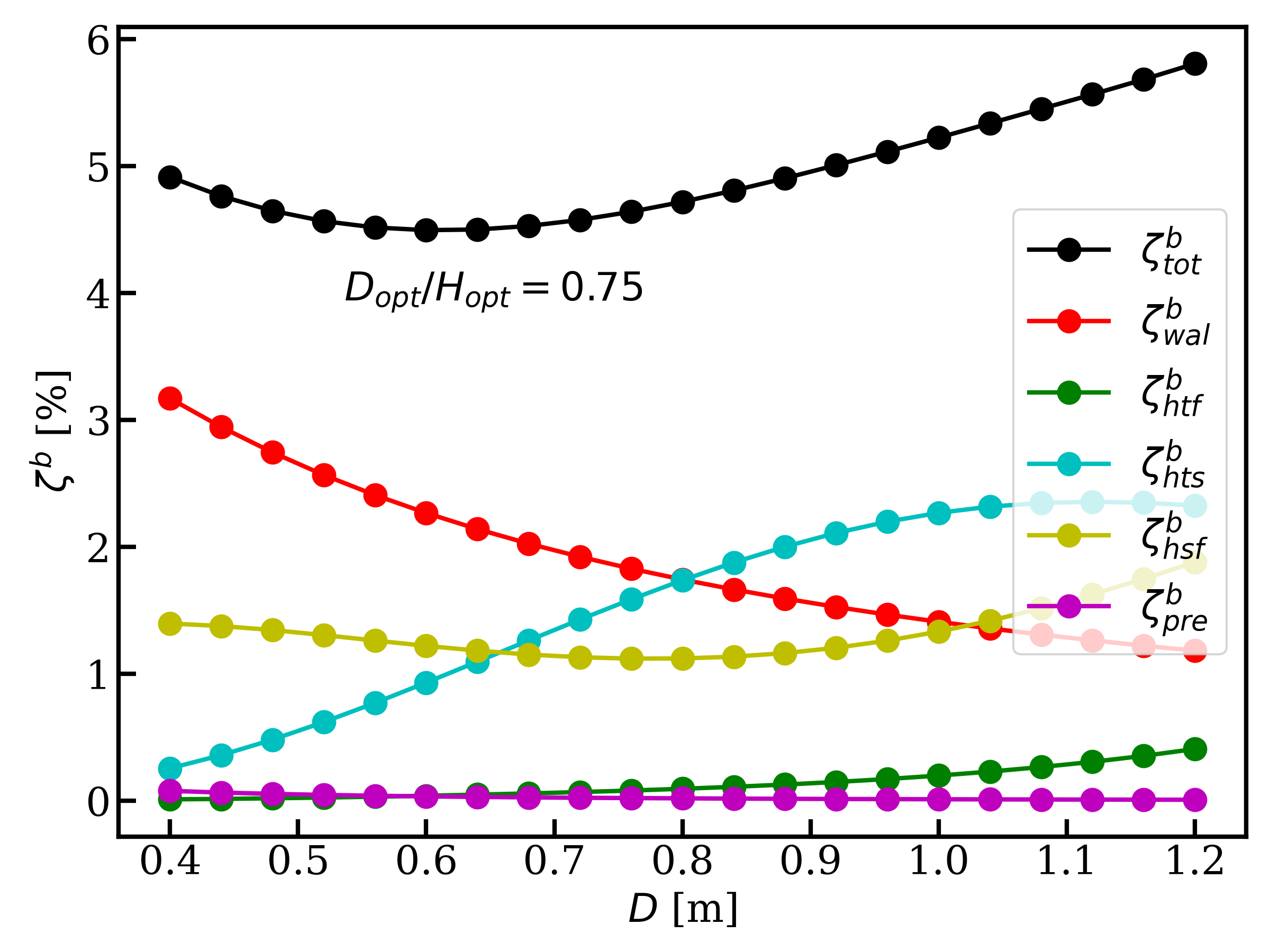}}
    \caption{Effect of the tank diameter $D$ on the exergy loss coefficients due to different mechanisms. The tank volume is kept constant ($0.23 \mathrm{m^3}$). The exit loss is not taken into account in the total loss.}
    \label{fig:loss_HD}
\end{figure}

The trend of the exergy loss coefficient $\zeta^b$ is similar to that of $\zeta^e$ since the exit loss dominates the total loss, see figure \ref{fig:loss_HD_exit}b. However, if the exit loss is not considered a real loss, then the other losses become significant, see figure \ref{fig:loss_HD}. Among these losses, the heat leakage loss coefficient $\zeta_{wal}^b$ decreases as $D$ increases, while the conduction loss coefficients $\zeta_{hts}^b$ and $\zeta_{htf}^b$ increase. The thermal loss coefficient $\zeta_{hsf}^b$ has a minimum value at $D=0.8\,\mathrm{m}$ and $H=0.45\,\mathrm{m}$. These losses lead to an optimal tank diameter of $D_{opt}=0.6\,\mathrm{m}$ and an optimal tank height of $H_{opt}=0.8\,\mathrm{m}$ ($D_{opt}/H_{opt}=0.75$). Through optimization, the total loss coefficient $\zeta^b_{tot}$ is minimized to $4.5\%$, whereas $\zeta^b_{tot}$ for the original storage tank is $4.9\%$.

\begin{figure}
    \centerline{\includegraphics[width=100mm]{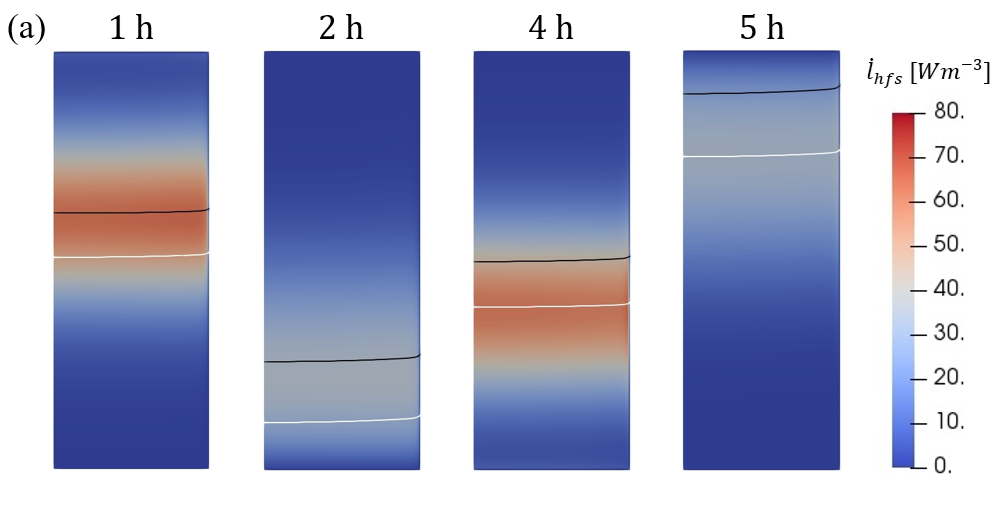}}
    \centerline{\includegraphics[width=100mm]{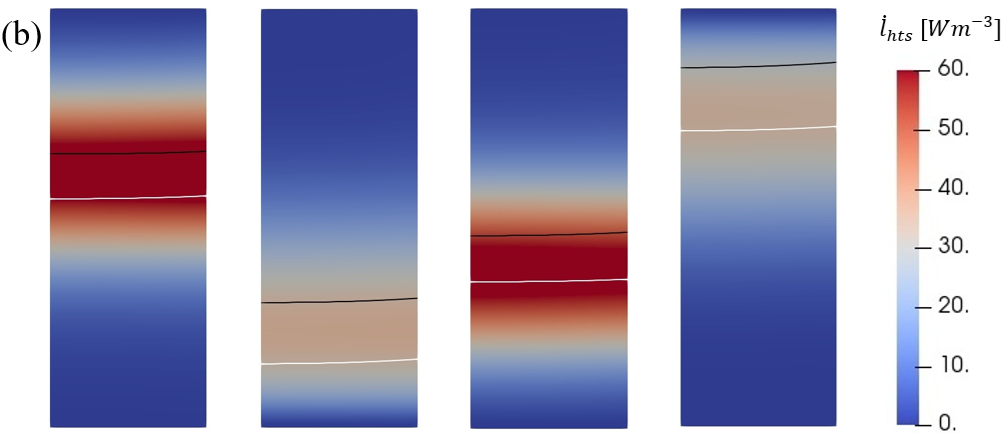}}
    \caption{Thermal loss intensity $\dot{l}_{hsf}$ (a) and solid conduction loss intensity $\dot{l}_{hts}$ during the charge ($0\mathrm{h}$-$3\mathrm{h}$) and discharge ($3\mathrm{h}$-$6\mathrm{h}$). The iso-surfaces of $T=465 \mathrm{K}$ (black lines) and $T=455 \mathrm{K}$ (white lines) are shown to indicate the location of the thermocline.}
    \label{fig:loss_HD_2d}
\end{figure}

Figure \ref{fig:loss_HD_2d} shows the locations and strengths of the thermal and solid conduction losses in the optimized storage tank. Compared to the original storage tank, the thermal loss intensity is weaker in the optimized tank, while the solid conduction loss becomes stronger, see figures \ref{fig:loss_Promes_2d} and \ref{fig:loss_HD_2d} for comparison. The thermocline still moves faster than the regions loss, however the lag is less evident. 

\subsection{Effect of particle size}
The particle size $d_p$ significantly affects energy or exergy losses. On the one hand, decreasing $d_p$ enhances the heat transfer of the fluid and solid. This reduces the exit and thermal losses, thus improving the efficiency of a TES system. However, a decrease in $d_p$ also causes a larger pressure loss, reducing the exergy storage efficiency $\eta_{II}$. Therefore, a SLA of the effect of $d_p$ on a SHS system often yields an optimal $d_p$, which results in a minimum total loss coeficient $\zeta^b_{tot}$. 

\begin{figure}
    \centerline{\includegraphics[width=80mm]{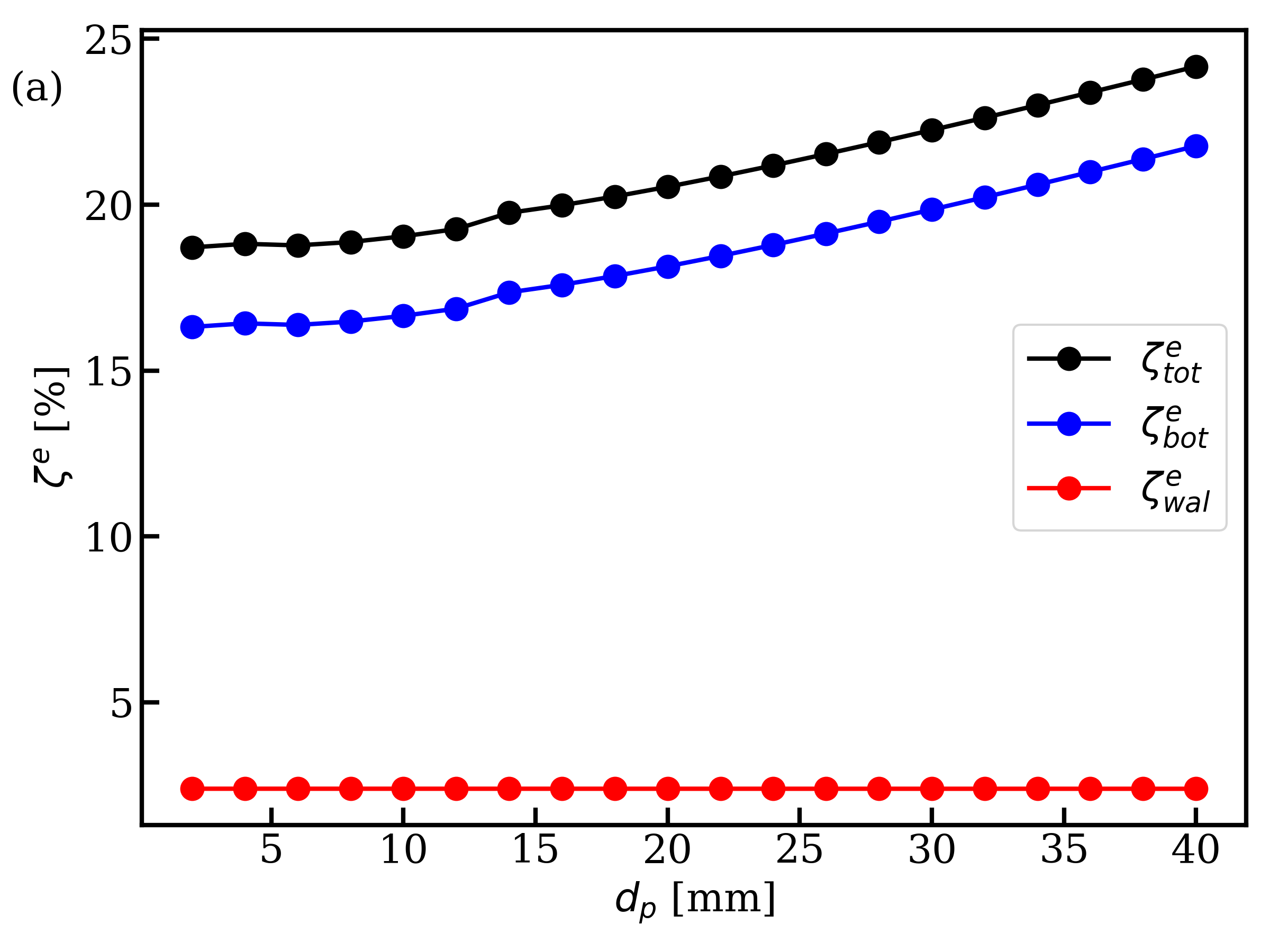}
                \includegraphics[width=80mm]{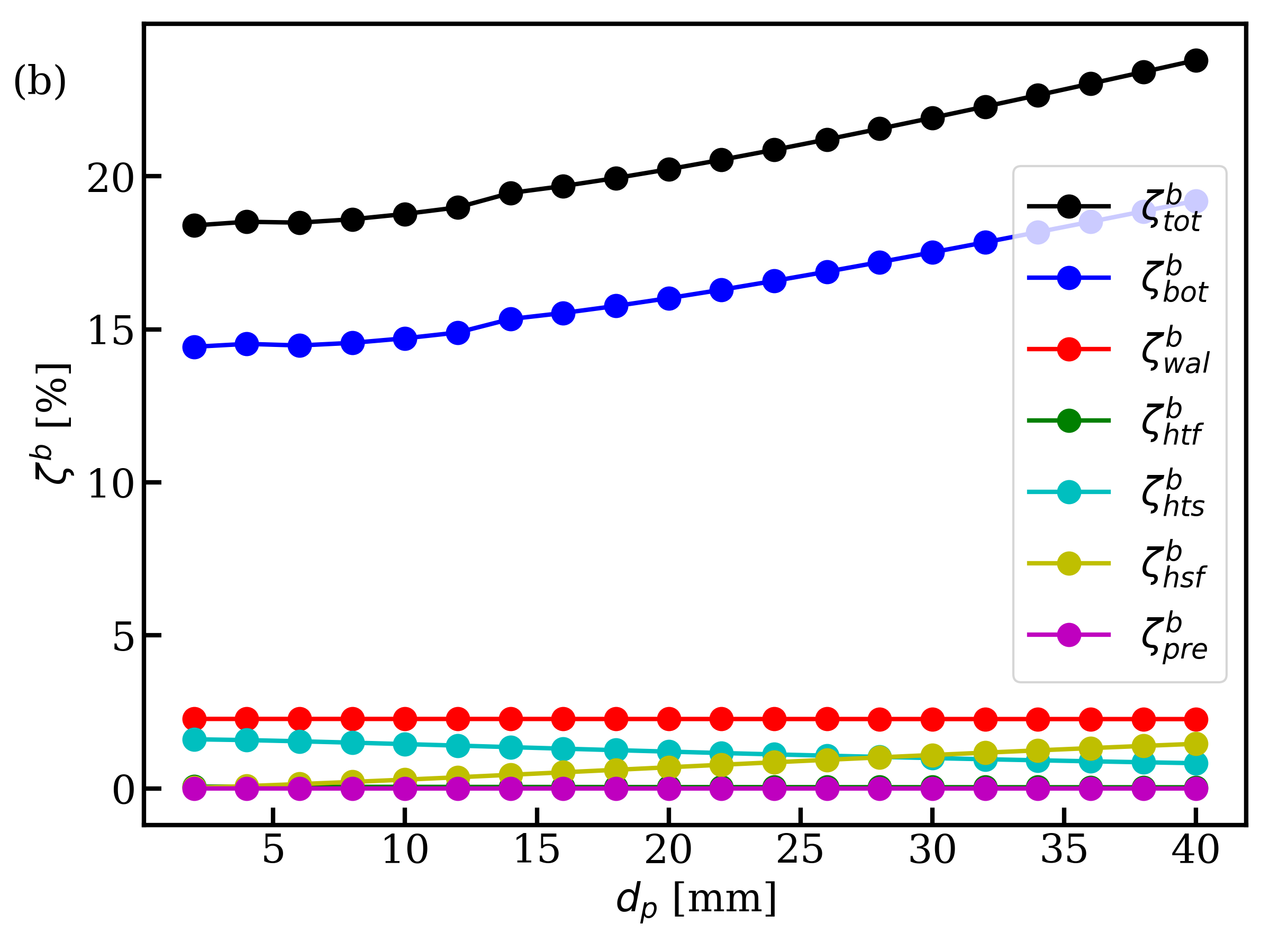}}
    \caption{Effect of the particle size $d_p$ on the energy (a) and exergy (b) loss coefficients due to different mechanisms. The tank hight and diameter are ($H=0.8 \, \mathrm{m}$) and ($D=0.6 \, \mathrm{m}$). The exit loss is taken into account in the total loss.}
    \label{fig:loss_dp_exit}
\end{figure}

However, the SLA of the present TES system does not provide an optimized $d_p$ value, see figure \ref{fig:loss_dp_exit}. The total energy loss coefficient $\zeta^e_{tot}$ and the exergy loss coefficient $\zeta^b_{tot}$ both decrease as $d_p$ decreases. This is because the pressure loss in the current storage tank is very small due to the low flow velocity ($v=8.4\times10^{-5} \, \mathrm{m/s}$ for the optimized tank). Using the particles with their original size ($d_p=40\,\mathrm{mm}$) results in a pressure drop of only $0.08\,\mathrm{pa}$, while reducing the particle size to $2\,\mathrm{mm}$ increases the pressure drop to $30.4\,\mathrm{pa}$, which is still not significantly high. The pressure loss is significantly smaller than the other losses. Consequently, we have not found an optimized $d_p$ hat minimizes the total exergy loss coefficient $\zeta^b_{tot}$. 

\begin{figure}
    \centerline{\includegraphics[width=100mm]{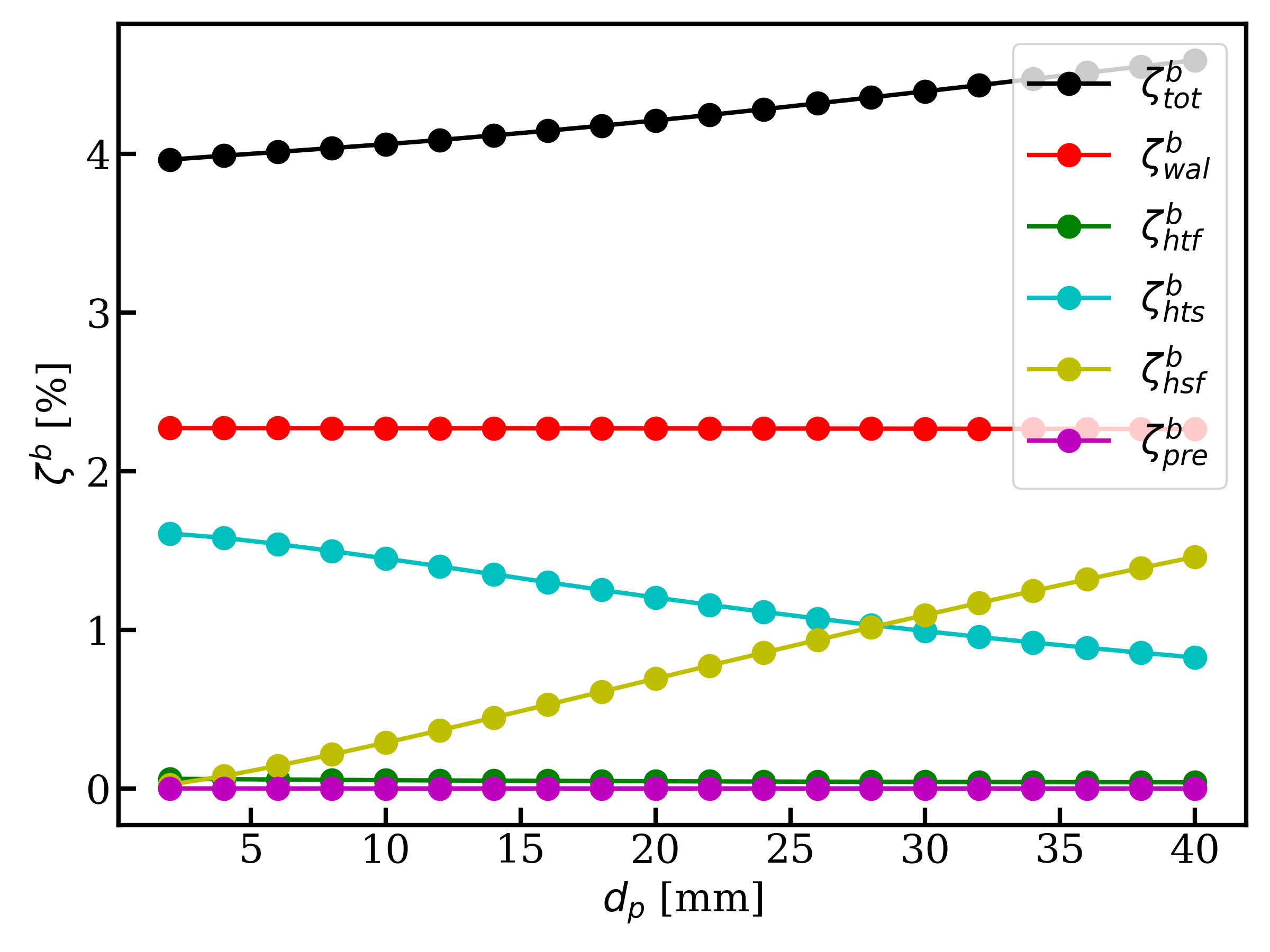}}
    \caption{Effect of the particle size $d_p$ on the exergy loss coefficients due to different mechanisms. The tank hight and diameter are ($H=0.8 \, \mathrm{m}$) and ($D=0.6 \, \mathrm{m}$). The exit loss is not taken into account in the total loss.}
    \label{fig:loss_dp}
\end{figure}

\begin{figure}
    \centerline{\includegraphics[width=100mm]{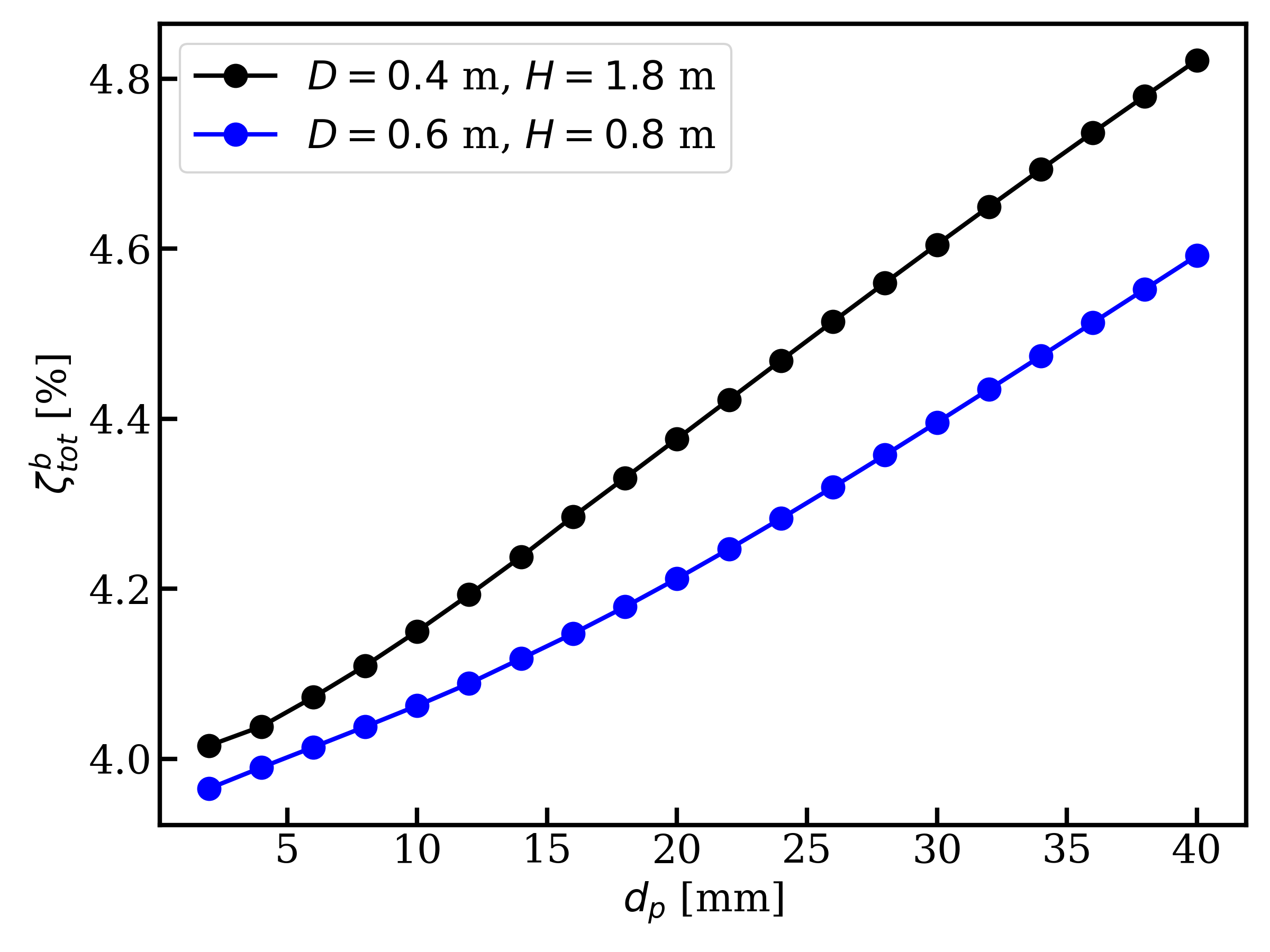}}
    \caption{Particle size $d_p$ versus total exergy loss coefficients for different tank hight-diameter ratios. The exit loss is not taken into account in the total loss.}
    \label{fig:loss_dp_DH}
\end{figure}

\begin{figure}
    \centerline{\includegraphics[width=100mm]{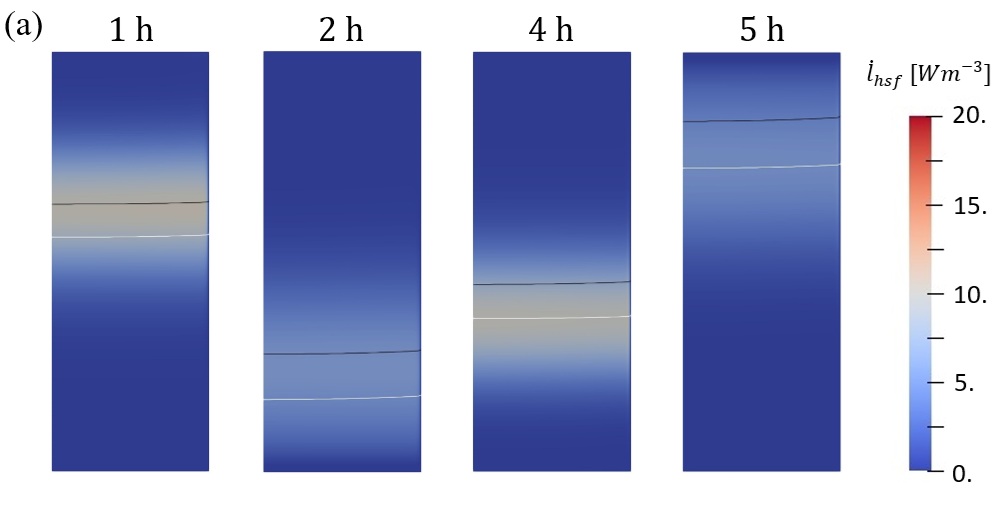}}
    \centerline{\includegraphics[width=100mm]{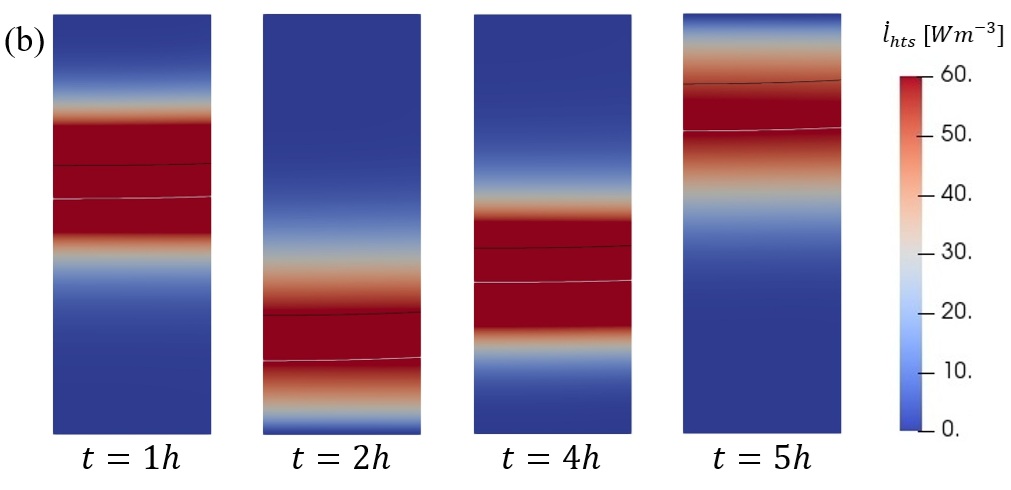}}
    \caption{Thermal loss intensity $\dot{l}_{hsf}$ (a) and solid conduction loss intensity $\dot{l}_{hts}$ during charge ($0\mathrm{h}$-$3\mathrm{h}$) and discharge ($3\mathrm{h}$-$6\mathrm{h}$). The iso-surfaces of $T=465 \mathrm{K}$ (black lines) and $T=455 \mathrm{K}$ (white lines) are shown to indicate the location of the thermocline.}
    \label{fig:loss_dp_2d}
\end{figure}

The total exergy loss coefficient $\zeta^b_{tot}$ does not provide an optimized particle size either when the exit loss is not considered, as shown in figure \ref{fig:loss_dp}. As the tank's aspect ratio $D/H$ decreases from 0.75 to 0.22 (the original value), the total exergy loss coefficient still decreases with a decrease in $d_p$, but more sharply (figure \ref{fig:loss_dp_DH}).  However, since considering $\zeta^b_{tot}$ decreases very slowly with a decrease in $d_p$ when $d_p\le8\,\mathrm{mm}$, we use the particles with $d_p=8\,\mathrm{mm}$ in our optimization. Figure \ref{fig:loss_dp_2d} shows that the thermal loss intensity weakens further as the particle size decreases, while the solid conduction loss intensity increases. Additionally, it can be seen that the thermocline becomes more parallel after the optimization, suggesting that the effect of the heat leakage at the wall becomes smaller.

\subsection{Truncated-cone-shaped storage tanks}
Based on the multi-dimensional CFD results, we can optimize the tank geometry to further improve the storage efficiency. One option for the shape optimization is replacing the current cylindrical tank with a truncated cone-shaped tank. The tank volume and hight are kept constant at $V=0.23\,\mathrm{m^3}$ and $H=0.8\,\mathrm{m}$.  The truncated cone has an upper surface diameter $D_1$ and a lower surface diameter $D_2$. We define an equivalent diameter $D_0$ as the diameter of a cylinder with the same volume and height as the truncated cone. For the cylinder optimized in section 4.2, the equivalent diameter is $D_0=0.8\,\mathrm{m}$. In this study, we have investigated $D_1$ in the range of $0.4\,\mathrm{m}$ to $1.0\,\mathrm{m}$ ($0.33 \le D_1/D_0 \le 1.67$).

\begin{figure}
    \centerline{\includegraphics[width=80mm]{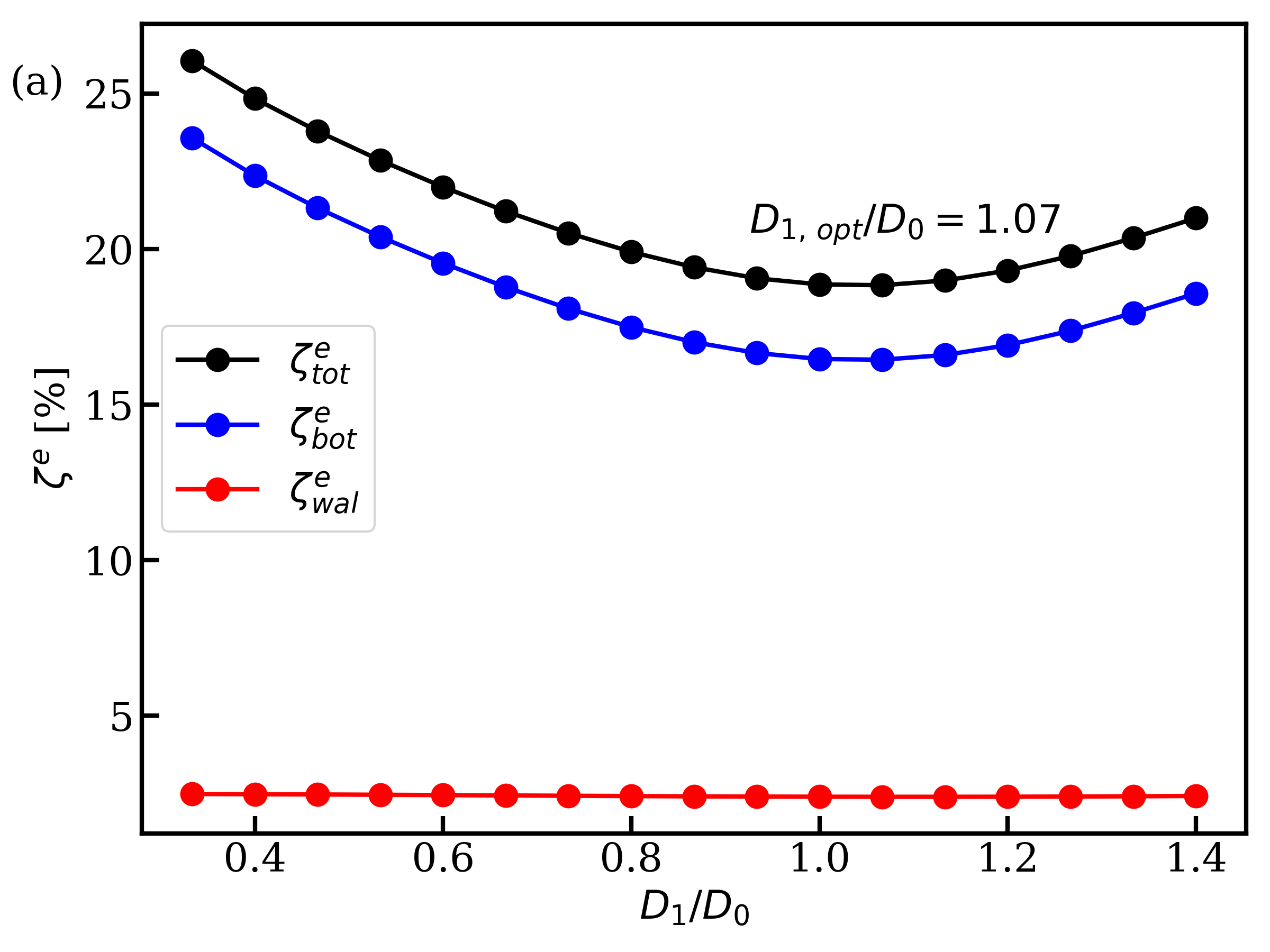}
                \includegraphics[width=80mm]{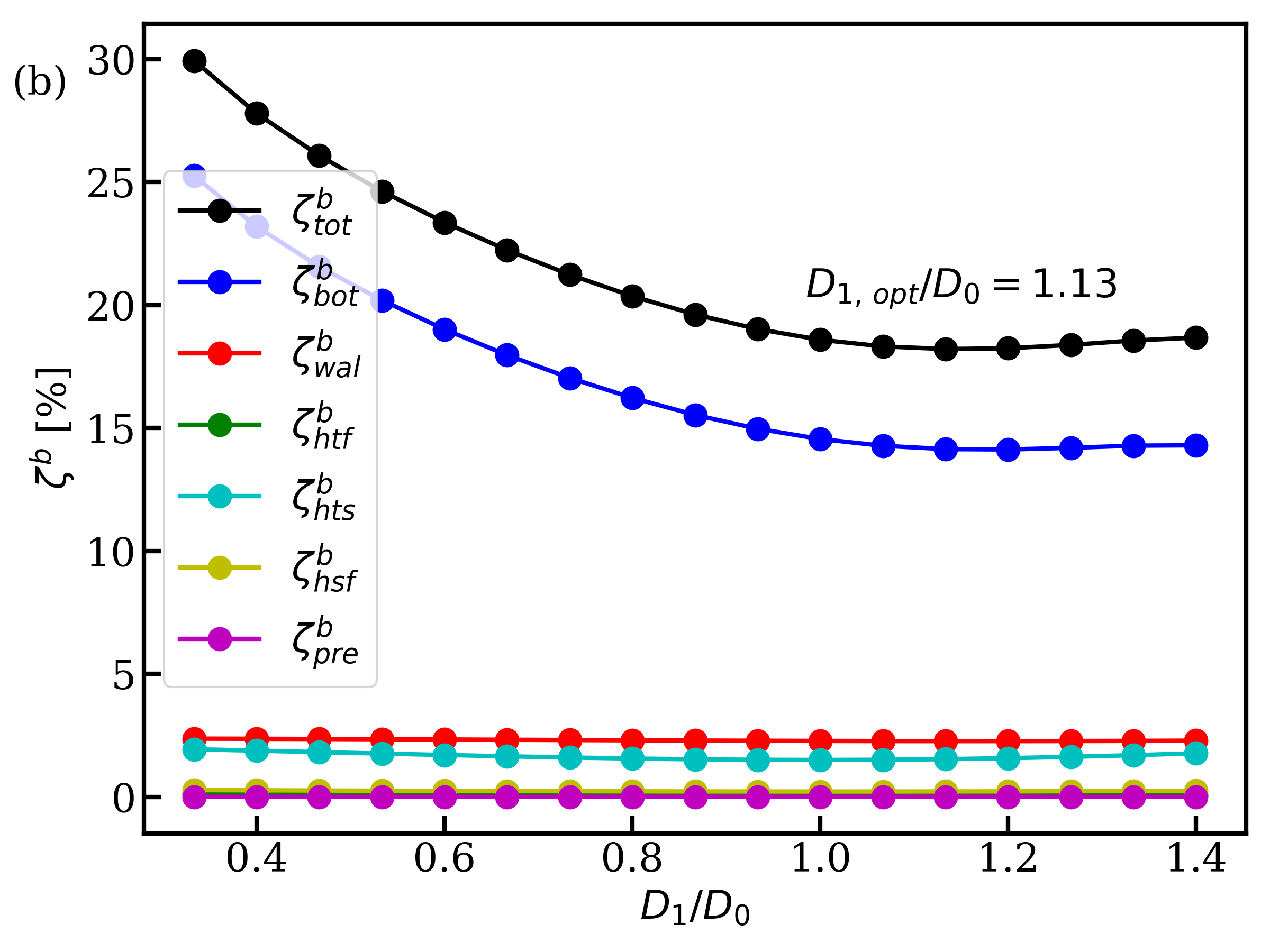}}
    \caption{Effect of the diameter ratio $D_1/D_0$ on the energy (a) and exergy (b) loss coefficients due to different mechanisms. The tank volume is kept as ($V=0.23 \, \mathrm{m}$). The exit loss is taken into account in the total loss.}
    \label{fig:loss_cone_exit}
\end{figure}

The effect of the diameter ratio $D_1/D_0$ on the energy and exergy loss coefficients is shown in figure \ref{fig:loss_cone_exit}. The exit loss is taken into account in the total loss in these figures. The CFD results show that $D_1$ should be slightly larger than $D_2$ to reduce the total energy or exergy loss coefficients. The minimum energy loss coefficient, $\zeta^e_{tot}=18.8\%$, is obtained as $D_1$ is set to the optimal value $D_{1,opt}=0.64\,m$ ($D_{1,opt}/D_0=1.07$). However, the SLA suggests a slightly higher optimal value of $D_{1,opt}=0.68\,m$ ($D_{1,opt}/D_0=1.13$). Additionally, $\zeta^b_{tot}$ increases very slowly as $D_1$ increases beyond its optimal value. The total energy and exergy exit loss coefficients at their optimal $D_1$ values are only slightly smaller than that of an equivalent cylindrical tank.

\begin{figure}
    \centerline{\includegraphics[width=100mm]{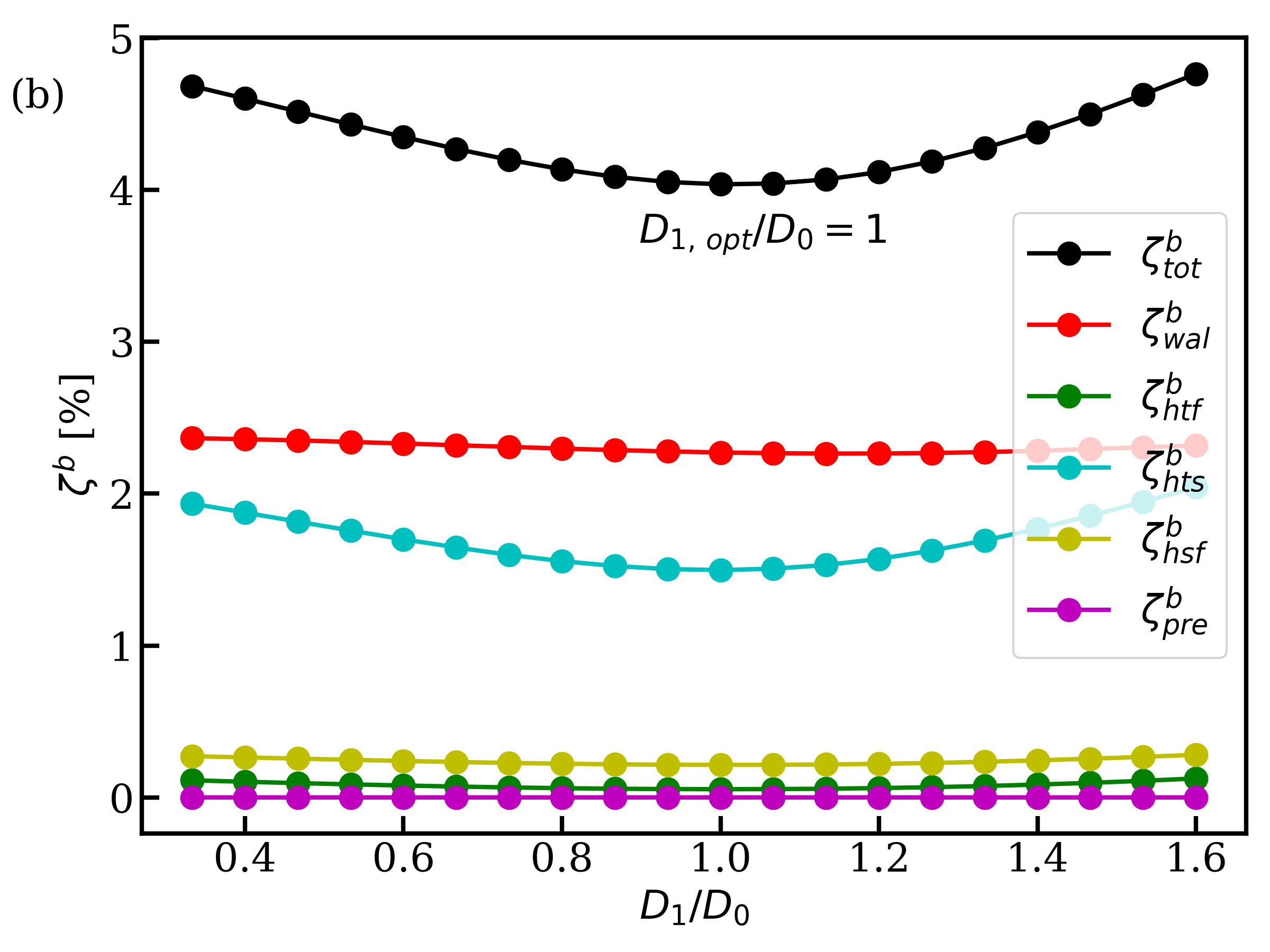}}
    \caption{Effect of the diameter ratio $D_1/D_0$ on the energy (a) and exergy (b) loss coefficients due to different mechanisms. The tank volume is kept as ($V=0.23 \, \mathrm{m}$). The exit loss is not taken into account in the total loss.}
    \label{fig:loss_cone}
\end{figure}

 If the exit loss is not considered a real loss, then the optimal value of $D_1$ is $D_{1,\,opt}=0.6\,m$ ($D_{1,\,opt})/D_0=1$), meaning the optimal geometry is a cylinder, see figure \ref{fig:loss_cone}. This differs from the study of Li, et al. (2018) \cite{Li2018}, who suggest that the truncated cone-shaped tank has better temperature stratification and thermal charging efficiency than cylindrical and spherical tanks. One possible reason of this discrepancy is that different HTF and THSM are used in this study, which may significantly affect on the optimal tank geometry. Additionally, a much more complicated heating system is studied in \cite{Li2018}, which has a significant effect on the losses. 

\begin{figure}
    \centerline{\includegraphics[width=100mm]{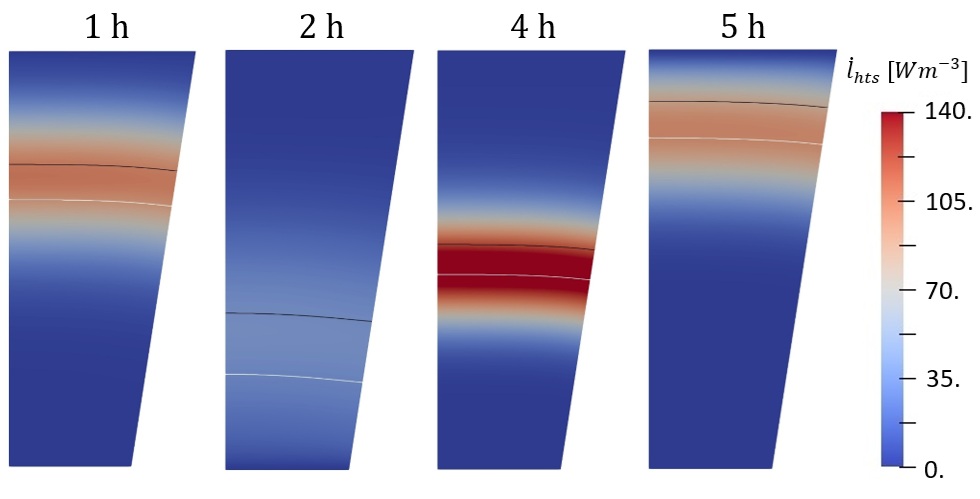}}
    \caption{Volume averaged exergy loss rates due to the heat transfer in the solid phase $\dot{l}_{hts}$ during charge ($0\mathrm{h}$-$3\mathrm{h}$) and discharge ($3\mathrm{h}$-$6\mathrm{h}$). The iso-surfaces of $T=465 \mathrm{K}$ (black lines) and $T=455 \mathrm{K}$ (white lines) are shown to indicate the location of the thermocline. The results for $D_{1}=0.68\, \mathrm{m}$, $D_{2}=0.48\, \mathrm{m}$, $H_{opt}=0.8\, \mathrm{m}$ and particle size ($d_{p,opt}=8\, \mathrm{mm}$) are shown.}
    \label{fig:loss_cone_2d}
\end{figure}

\begin{figure}
    \centerline{\includegraphics[width=80mm]{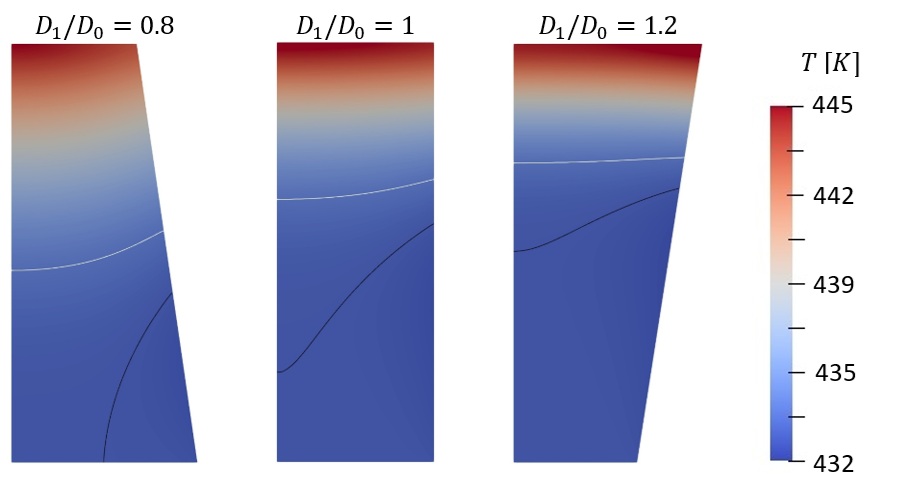}}
    \caption{Temperature fields for $D_1/D_0=0.8$, $1$ and $1.2$ at the end of discharge ($t=6\,\mathrm{h}$). The white and black lines indicate the iso-surfaces of $T=434K$ and $433K$, respectively.}
    \label{fig:temperature_cone}
\end{figure}

\begin{figure}
    \centerline{\includegraphics[width=80mm]{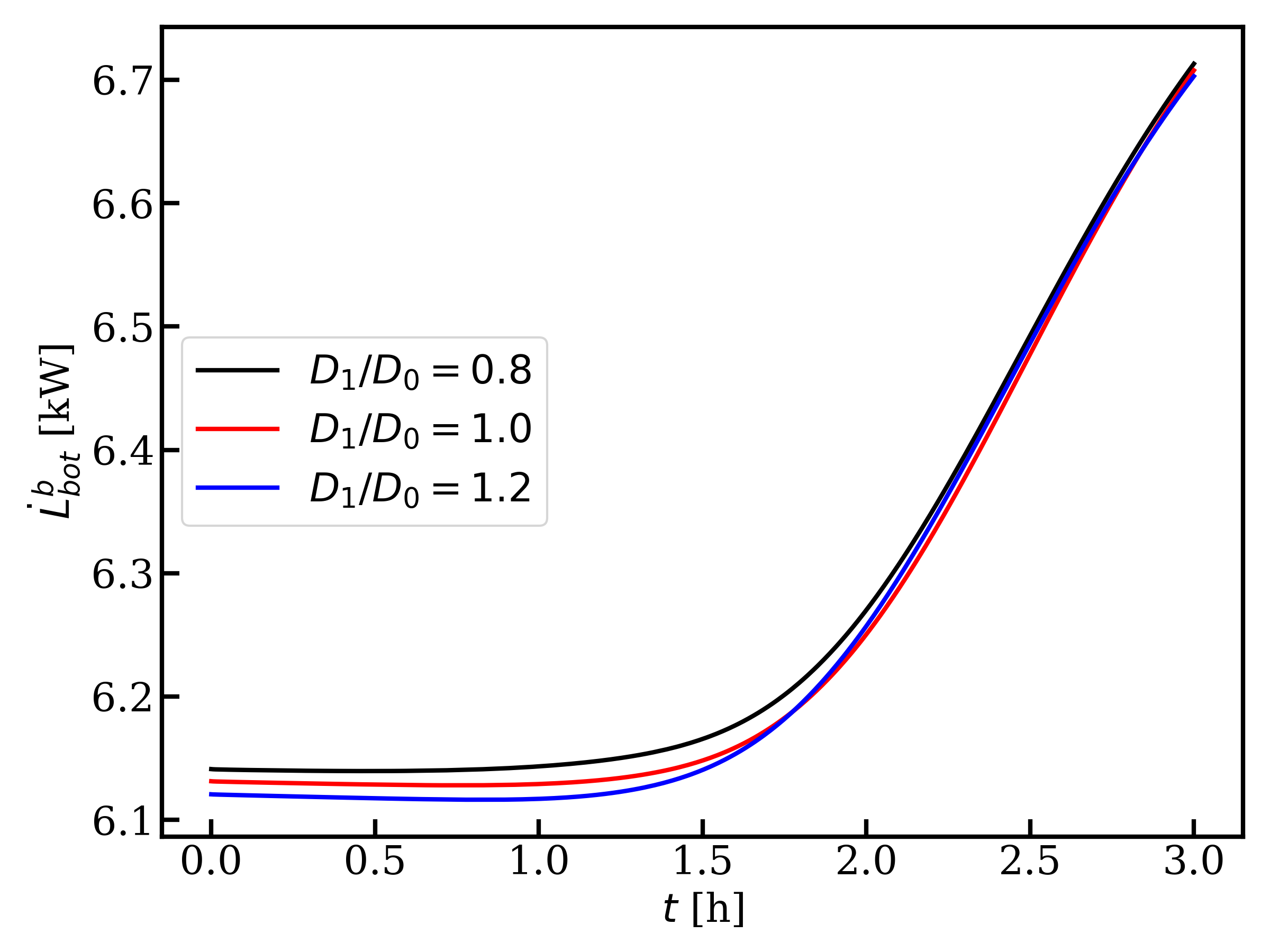}}
    \caption{Time evolution of the exit exergy loss rate $\dot{L}^b_{bot}$ for $D_1/D_0=0.8$, $1$ and $1.2$ during the charge.}
    \label{fig:loss_cone_charge}
\end{figure}

Figure \ref{fig:loss_cone_2d} shows the solid conduction loss intensity $\dot{l}_{hts}$ at typical times during a charge-discharge cycle. The change in $\dot{l}_{hts}$ is the most significant among internal losses when the exit loss is not considered, see figure \ref{fig:loss_cone}. As can be seen, the solid conduction loss during discharge is much stronger than during charge. This is due to the strong thermal conduction in the solid phase at the start of of the discharge process. Another interesting phenomenon is that, as $D_1/D_0$ increases, the total energy loss coefficient $\zeta^e_{tot}$ increases, however, $\zeta^b_{tot}$ almost does not change. Figure \ref{fig:temperature_cone} shows the temperature fields at the end of the discharge process (or the start of the charge process) for three $D_1/D_0$ values ($0.8$, $1.0$ and $1.2$) and explains this phenomenon. On the one hand, the remaining energy in the tank after the discharge increases as $D_1/D_0$ increases. This results in a higher energy loss coefficient $\zeta^e$ during charging. However, the iso-surfaces of $T=433\,K$ and $434\,K$ shown in the figure suggest that, as $D_1/D_0$ increases, the temperature becomes more uniformly distributed near the tank bottom. 
This leads to higher entropy near the tank bottom, which reduces the exergy leaving the tank during charging. Figure \ref{fig:loss_cone_charge} shows the time evolution of the exit loss rate $\dot{L}^b_{bot}$ during the charge. It confirms that, as $D_1/D_0$ increases, $\dot{L}^b_{bot}$ is lower during the earlier discharge period and higher during the later period. This explains why $\zeta^b_{tot}$ changes very slightly as $D_1/D_0$ increases beyond its optimal value. 

\subsection{Final results of optimization}
Summarizing the results above, we have proposed an optimized packed-bed SHS system. Based on the main parameters shown in table \ref{tab:Promes}, we recommend using a truncated cone-shaped tank with height of $H_{opt}=0.8\,\mathrm{m}$, an upper surface diameter of $D_{1,\,opt}=0.68\,\mathrm{m}$ and a lower surface diameter of $D_{2,\,opt}=0.52\,\mathrm{m}$ ($D_{1,\,opt}/D_0=1.13$). The tank is filled with the particles of $d_{p,\,opt}=8\,\mathrm{mm}$, while the other parameters remain unchanged. Through optimization, the total exergy loss coefficient $\zeta^b_{tot}$ is reduced from $4.9\%$ to $4.1\%$ when the exit loss is not considered a real loss. If the exit loss is considered, the energy loss coefficient $\zeta^e_{tot}$ is slightly reduced from $19.7\%$ to $19.0\%$, while $\zeta^b_{tot}$ is almost unchanged ($18.6\%$). As the exit loss is often not a real loss, the improvement in storage efficiency is apparent.

\section{Conclusions}
In this study, we developed macroscopic entropy and exergy transport equations for fluid flow and heat transfer in porous media. These equations allow us to perform a SLA analysis of the charge-discharge process of a SHS system using CFD results. Based on the SLA, we can determine where exergy is destroyed and how strong it is. This allows us to find ways to reduce losses and improve storage efficiency.

Using the laboratory scale packed-bed SHS system developed in the PROMES-CNRS laboratory as a test case, we have demonstrated how to perform a SLA with CFD results and optimize the design.  The CFD results for this system show that energy loss primarily occurs at the tank bottom (exit loss) during charge and on the wall surface (heat leakage loss). In addition to the exit loss and the heat leakage loss, the SLA shows that thermal loss (due to heat transfer between the fluid and solid) and solid conduction loss (due to thermal conduction in the solid) are also significant contributors to the total exergy loss, while the other losses are negligible for the studied SHS system. Unlike the exit and heat leakage losses, which occur at the boundary surfaces, the thermal and solid conduction losses occur inside the tank, close to the thermocline. The thermocline moves faster than the regions where these losses occur. This results in a delay in the emergence of the losses.

The SLA suggests an optimized tank aspect ratio of $D_{opt}/H_{opt}=0.75$, which yields a minimum total exergy loss coefficient $\zeta^b_{tot}$ as the exit loss is not considered as a real loss. As the exit loss is considered, both energy and exergy loss coefficients $\zeta^e_{tot}$ and $\zeta^b_{tot}$ decrease as $D/H$ decreases, since this reduces the exit loss, which dominates the total loss. Additionally, both $\zeta^e_{tot}$ and $\zeta^b_{tot}$ decrease with $d_p$, regardless of whether the exit loss is considered. The pressure loss is negligible for the SHS system studied. The SLA recommends a tank with a truncated cone shape, with a slightly larger the upper surface ($D_{1,\,opt}/D_0=1.13$). This tank geometry yields the lowest value of $\zeta^b_{tot}$ as the exit loss is considered. Through the SLA, we have proposed a SHS system with a tank aspect ratio of $D_{opt}/H_{opt}=0.75$, a diameter ratio of $D_{1,opt}/D_0=1.13$ and a particle size $d_{p,opt}=8\,\mathrm{mm}$. Without considering the exit loss, the total exergy loss coefficient $\zeta^b_{tot}$ is reduced from $4.9\%$ for the original design to $4.1\%$ for the optimized design. The study also shows that comprehensive consideration of both energy and exergy efficiencies should be given during the design process. The developed entropy and exergy transport equations will be applied to more complicated energy storage systems in the future work for further validation.

\renewcommand{\nomgroup}[1]{%
  \ifthenelse{\equal{#1}{A}}{\item[\textbf{Roman letters}]}{%
  \ifthenelse{\equal{#1}{B}}{\item[\textbf{Greek symbols}]}{%
  \ifthenelse{\equal{#1}{D}}{\item[\textbf{Subscripts}]}{%
  \ifthenelse{\equal{#1}{C}}{\item[\textbf{Superscripts}]}{%
  \ifthenelse{\equal{#1}{E}}{\item[\textbf{Abbreviations}]}{}}}}}}

\nomenclature[A]{$B$}{Exergy, $\mathrm{J}$} 
\nomenclature[A]{$b$}{Specific exergy, $\mathrm{J\,kg^{-1}}$} 
\nomenclature[A]{$e$}{Specific energy, $\mathrm{J\,kg^{-1}}$} 
\nomenclature[A]{$s$}{Specific entropy, $\mathrm{J\,kg^{-1}\,T^{-1}}$} 
\nomenclature[A]{$\dot{l}$}{Loss rate intensity, $\mathrm{W\,m^{-3}}$}
\nomenclature[A]{$\dot{L}$}{Loss rate, $\mathrm{W}$}
\nomenclature[A]{$t$}{Time, s}
\nomenclature[A]{$T$}{Temperature, K}
\nomenclature[A]{$C_p$}{Heat capacity, $\mathrm{J\,kg^{-1}\, K^{-1}}$}
\nomenclature[A]{$k$}{Heat conductivity, $\mathrm{J\,m^{-1}K^{-1}}$}
\nomenclature[A]{$Q$}{Thermal energy, $\mathrm{J}$}
\nomenclature[A]{$d_p$}{Particle size, $\mathrm{m}$}
\nomenclature[A]{$K$}{Permeability, $\mathrm{m^2}$}
\nomenclature[A]{$H$}{Tank height, $\mathrm{m}$}
\nomenclature[A]{$D$}{Tank diameter, $\mathrm{m}$}
\nomenclature[A]{$D_0$}{Equivalent tank diameter, $\mathrm{m}$}
\nomenclature[A]{$D_1$}{Upper surface diameter of a truncated-cone shaped tank, $\mathrm{m}$}
\nomenclature[A]{$D_2$}{Lower surface diameter of a truncated-cone shaped tank, $\mathrm{m}$}

\nomenclature[B]{$\phi$}{Porosity, -}
\nomenclature[B]{$\mu$}{Dynamic viscosity, Pa s}
\nomenclature[B]{$\rho$}{Density, $\mathrm{kg\,m^{-3}}$}
\nomenclature[B]{$\zeta$}{Loss coefficient, -}
\nomenclature[B]{$\eta$}{Storage efficiency, -}

\nomenclature[C]{$e$}{Energy}
\nomenclature[C]{$b$}{Exergy}
\nomenclature[C]{$mic$}{Microscopic}
\nomenclature[C]{$mac$}{Macroscopic}

\nomenclature[D]{$cyc$}{Charge-discharge cycle}
\nomenclature[D]{$vol$}{Tank volume}
\nomenclature[D]{$s$}{Solid}
\nomenclature[D]{$f$}{Fluid}
\nomenclature[D]{$tot$}{Total}
\nomenclature[D]{$w$, $wal$}{Tank wall}
\nomenclature[D]{$bot$}{Tank bottom surface}
\nomenclature[D]{$hsf$}{Heat transfer between fluid and solid}
\nomenclature[D]{$hts$}{Heat conduction in solid}
\nomenclature[D]{$htf$}{Heat conduction in fluid}
\nomenclature[D]{$pre$}{Pressure}
\nomenclature[D]{$oth$}{Others}
\nomenclature[D]{$dis$}{Discharge}
\nomenclature[D]{$opt$}{Optimized}
\nomenclature[D]{$chg$}{Charge}
\nomenclature[D]{$I$}{The first law}
\nomenclature[D]{$II$}{The second law}

\nomenclature[E]{TES}{Thermal energy storage}
\nomenclature[E]{TCHS}{Thermo-chemical heat storage}
\nomenclature[E]{TESM}{Thermal energy storage material}
\nomenclature[E]{HTF}{Heat transfer fluid}
\nomenclature[E]{SHS}{Sensible heat storage}
\nomenclature[E]{SLA}{Second law analysis}
\nomenclature[E]{CFD}{Computational fluid dynamics}
\printnomenclature


\section*{Acknowledgments}
The authors gratefully acknowledge the support of this study by the DFG (Deutsche Forschungsgemeinschaft, 552151258) and the computing center of Hamburg University of Technology (RZ-TUHH).

\section*{Declaration of interests}
The authors report no conflict of interest.

\section*{Data availability statement}
The data are available through direct contact with the author.

\bibliographystyle{unsrt}  
\bibliography{references}

\end{document}